\documentclass[final]{IEEEtran}
\usepackage{latexsym}
\usepackage{amsfonts}
\usepackage{cite}
\usepackage{amssymb}
\usepackage{bm}
\usepackage{amsmath}
\usepackage{amssymb}
\usepackage{graphicx}
\usepackage{mathrsfs}
\usepackage{epsfig}
\usepackage{psfrag}
\usepackage{setspace}
\usepackage{hyperref}
\usepackage{algorithm}
\usepackage{algpseudocode}
\usepackage{stfloats}

\usepackage{theorem}
\theoremheaderfont{\bfseries\upshape} \theoremstyle{plain}

\theorembodyfont{\rmfamily}

\theoremheaderfont{\itshape}
\newtheorem{remark}{Remark}

\theorembodyfont{\itshape}
\newtheorem{theorem}{Theorem}
\newtheorem{proposition}{Proposition}
\newtheorem{lemma}{Lemma}

\theorembodyfont{\rmfamily}
\newtheorem{definition}{Definition}

\def \cX {\mathcal{X}}
\def \cY {\mathcal{Y}}

\def \bX {\mathbf{X}}
\def \bY {\mathbf{Y}}

\def\mathllap{\mathpalette\mathllapinternal}
\def\mathllapinternal#1#2{\llap{$\mathsurround=0pt#1{#2}$}}
\DeclareRobustCommand\rH{{H\mathllap{\overline{\vphantom{H}\hphantom{\rule{7.5pt}{0pt}}}\mspace{0.7mu}}}}
\DeclareRobustCommand{\rI}{I\mathllap{\overline{\vphantom{I}\hphantom{\rule{3pt}{0pt}}}\mspace{0.7mu}}}

\newcommand{\PR}{\mathbb{P}} 
\newcommand{\E}{\mathbb{E}} 
\newcommand{\norm}[1]           {\| #1\|}
\newcommand{\funcf}[1]           {f( #1)}
\newcommand{\DI}[2] {I( #1 \to #2)}
\newcommand{\DIR}[2] {\rI( #1 \to #2)}

\title{Universal Estimation of Directed Information}

\author{Jiantao~Jiao,~\IEEEmembership{Student Member,~IEEE},~Haim~H.~Permuter,~\IEEEmembership{Member,~IEEE},~Lei~Zhao,~Young-Han~Kim,~\IEEEmembership{Senior Member,~IEEE}, and Tsachy~Weissman,~\IEEEmembership{Fellow,~IEEE}
\thanks{Manuscript received Month 00, 0000; revised Month 00, 0000; accepted Month 00, 0000. Date of current version Month 00, 0000.
This work was supported in part by the Center for Science of Information (CSoI), an NSF Science and Technology Center, under grant agreement CCF-0939370, the US--Israel Binational Science Foundation (BSF) Grant 2008402, NSF Grant CCF-0939370, and the Air Force Office of Scientific Research (AFOSR) through Grant FA9550-10-1-0124. Haim H. Permuter was supported in part by the Marie Curie Reintegration Fellowship.
The material in this paper was presented in part at the 2010 IEEE International Symposium on
Information Theory, Austin, TX, and the 2012 IEEE International Symposium on
Information Theory, Cambridge, MA.}%
\thanks{Jiantao Jiao is with the Department of Electrical Engineering, Stanford University, Stanford, CA 94305, USA (e-mail: {jiantao@stanford.edu}).}%
\thanks{Haim Permuter is with the Department of Electrical and Computer Engineering, Ben-Gurion University of the Negev, Beer-Sheva 84105, Israel (e-mail: {haimp@bgu.ac.il}).}%
\thanks{Lei Zhao was with the Department of Electrical Engineering, Stanford University, Stanford CA, USA. He is now with Jump Operations, Chicago, IL 60654, USA (e-mail: zhaolei122@gmail.com).}%
\thanks{Young-Han Kim is with the Department of Electrical and Computer Engineering, University of California, San Diego, La Jolla, CA 92093, USA (e-mail: {yhk@ucsd.edu}).}%
\thanks{Tsachy Weissman is with the Department of Electrical Engineering, Stanford University, Stanford CA 94305, USA (e-mail: tsachy@stanford.edu).}
\thanks{Communicated by I. Kontoyiannis, Associate Editor for Shannon Theory.}%
\thanks{Color versions of one or more of the figures in this paper are available
online at http://ieeexplore.ieee.org.}%
\thanks{Digital Object Identifier 10.1109/TIT.2013.0000000}%
}

\begin{document}

\maketitle

\begin{abstract}
Four estimators of the directed information rate between a pair of jointly stationary ergodic finite-alphabet processes are proposed, based on universal probability assignments. The first one is a Shannon--McMillan--Breiman type estimator, similar to those used by Verd\'u (2005) and Cai, Kulkarni, and Verd\'u (2006) for estimation of other information measures. We show the almost sure and $L_1$ convergence properties of the estimator for any underlying universal probability assignment. The other three estimators map universal probability assignments to different functionals, each exhibiting relative merits such as smoothness, nonnegativity, and boundedness. We establish the consistency of these estimators in almost sure and $L_1$ senses, and derive near-optimal rates of convergence in the minimax sense under mild conditions.  These estimators carry over directly to estimating other information measures of stationary ergodic finite-alphabet processes, such as entropy rate and mutual information rate,  with near-optimal  performance and provide alternatives  to classical approaches in the existing literature. Guided by these theoretical results, the proposed estimators are implemented using the context-tree weighting algorithm as the universal probability assignment. Experiments on synthetic and real data are presented, demonstrating the potential of the proposed schemes in practice and the utility of directed information estimation in detecting and measuring causal influence and delay.
\end{abstract}


\newpage

\section{Introduction}
\IEEEPARstart{F}{irst} introduced by Marko \cite{Marko1973} and Massey \cite{Massey90}, directed information arises as a natural counterpart of mutual information for channel
capacity when causal feedback from the receiver to the sender is present. In \cite{Kramer1998} and
\cite{Kramer03}, Kramer extended the use of directed information to
discrete memoryless networks with feedback, including the two-way
channel and the multiple access channel. Tatikonda and
Mitter~\cite{Tatikonda--Sanjoy2009} used  directed
information spectrum  to establish a general feedback channel coding theorem for
channels with memory. For a class of stationary
channels with feedback, where the output is a function of the current
and past $m$ inputs and channel noise, Kim \cite{Kim08} proved that
the feedback capacity is equal to the limit of the maximum
normalized directed information from the input to the
output.  Permuter, Weissman, and Goldsmith~\cite{Permuter--Weissman--Goldsmith2009} considered the capacity of discrete-time finite-state channels with
feedback where the feedback is a time-invariant
function of the output. Under mild conditions, they showed that the
capacity is again the limit of the maximum normalized directed information.
Recently,
Permuter, Kim, and Weissman \cite{Permuter--Kim--Weissman2009} showed that directed
information plays an important role in portfolio theory, data
compression, and hypothesis testing under causality constraints.

Beyond information theory, directed information is a valuable tool in biology, for it provides an alternative to the notion of Granger causality\cite{Granger1969}, which has been perhaps the most widely-used means of identifying causal influence between two random processes. For example, Mathai, Martins, and Shapiro\cite{Mathai--Martins--Shapiro2007} used directed information to identify pairwise influence in gene networks. Similarly, Rao, Hero, States, and Engel\cite{Rao--Hero--States--Engel2008} used directed information to test the direction of influence in gene networks.

Since directed information has significance in various fields, it is of both theoretical and practical importance to develop efficient methods of estimating it. The problem of estimating information measures, such as entropy, relative entropy and mutual information, has been extensively studied in the literature.
Verd\'u \cite{Verdu2005} gave an overview of universal estimation of information measures. Wyner and Ziv \cite{Wyner--Ziv1989} applied the idea of Lempel--Ziv parsing to estimate entropy rate, which converges in probability for all stationary ergodic processes. Ziv and Merhav \cite{Ziv--Merhav1993} used Lempel--Ziv parsing to estimate relative entropy (Kullback--Leibler divergence) and established consistency under the assumption that the observations are generated by independent Markov sources. Cai, Kulkarni, and Verd\'u \cite{Cai--Kulkarni--Verdu2006} proposed two universal relative entropy estimators for finite-alphabet sources, one based on the Burrows--Wheeler
transform (BWT)~\cite{Burrows--Wheeler1994} and the other based on the context-tree weighting (CTW) algorithm \cite{Willems--Shtarkov--Tjalkens1995}. The BWT-based estimator was applied in universal entropy estimation by Cai, Kulkarni, and Verd\'u \cite{Cai--Kulkarni--Verdu2004}, while the CTW-based one was applied in universal erasure entropy estimation by Yu and Verd\'u \cite{Yu--Verdu2006}.

  For the problem of estimating directed information, Quinn, Coleman, Kiyavashi, and Hatspoulous \cite{Quinn--Coleman--Kiyavash--Hatsopoulos2009} developed an estimator to infer causality in an ensemble of neural spike train recordings. Assuming
  a parametric generalized linear model and stationary ergodic Markov processes, they established strong consistency results. Compared to \cite{Quinn--Coleman--Kiyavash--Hatsopoulos2009}, Zhao, Kim, Permuter, and Weissman\cite{Zhao--Kim--Permuter--Weissman2010} focused on universal methods for arbitrary stationary ergodic processes with finite alphabet and showed their $L_1$ consistencies.

  As an improvement and extension of \cite{Zhao--Kim--Permuter--Weissman2010}, the main contribution of this paper is a general framework for estimating information measures of stationary ergodic finite-alphabet processes, using ``single-letter'' information-theoretic functionals. Although our methods can be applied in estimating a number of information measures, for concreteness and relevance to emerging applications we focus on estimating the directed information rate between a pair of jointly stationary ergodic finite-alphabet processes.

  The first proposed estimator is adapted from the universal relative entropy estimator in \cite{Cai--Kulkarni--Verdu2006} using the CTW algorithm, and we provide a refined analysis yielding strong consistency results. We further propose three additional estimators in a unified framework, present both weak and strong consistency results, and establish near-optimal rates of convergence under mild conditions. We then employ our estimators on both simulated and real data, showing their effectiveness in measuring channel delays and causal influences between different processes. In particular, we use these estimators on the daily stock market data from 1990 to 2011 to observe a significant level of causal influence from the Dow Jones Industrial Average to the Hang Seng Index, but relatively low causal influence in the reverse direction.

The rest of the paper is organized as follows. Section~II reviews preliminaries on directed information, universal probability assignments, and the context-tree weighting algorithm.
Section~III presents our proposed estimators and their basic properties. Section~IV is dedicated to performance guarantees of the proposed estimators, including their consistencies and miminax-optimal rates of convergence.
Section V shows experimental results in which we apply the proposed estimators to simulated and real data. Section~VI concludes the paper. The proofs of the main results are given in the Appendices.

\section{Preliminaries}
We begin with mathematical definitions of directed information and causally conditional entropy.
We also define universal and pointwise universal probability assignments. We then introduce the context-tree weighting (CTW) algorithm used in our implementations of the universal estimators that are introduced in the next section.

Throughout the paper, we use uppercase letters $X, Y, \ldots$ to denote random variables and lowercase letters $x, y, \ldots$ to denote values they assume.
By convention, $X = \emptyset$ means that $X$ is a degenerate random variable (unspecified constant) regardless of its support.
We denote the $n$-tuple $(X_1,X_2,\ldots,X_n)$ as $X^n$ and $(x_1,x_2,\ldots,x_n)$ as $x^n$.
Calligraphic letters $\cX, \cY, \ldots$ denote alphabets of $X, Y, \ldots$,
and $|\cX|$ denotes the cardinality of $\cX$. Boldface letters $\bX, \bY, \ldots$ denote stochastic processes, and throughout this paper, they are finite-alphabet.

Given a probability law $P$, $P(x^i) = P\{X^i = x^i\}$
denotes the probability mass function (pmf) of $X^i$ and $P(x_i | x^{i-1})$ denotes the conditional
pmf of $X_i$ given $\{X^{i-1} = x^{i-1}\}$, i.e., with slight abuse of notation, $x_i$ here is a dummy variable and $P(x_i|x^{i-1})$ is an element of $\mathcal{M}(\mathcal{X})$, the probability simplex on $\mathcal{X}$, representing the said conditional pmf. Accordingly, $P(x_i | X^{i-1})$ denotes the conditional pmf $P(x_i|x^{i-1})$ evaluated for the random sequence $X^{i-1}$, which is an $\mathcal{M}(\mathcal{X})$-valued random vector, while $P(X_i | X^{i-1})$ is the random variable denoting the $X_i$-th component of $P(x_i|X^{i-1})$. Throughout this paper, $\log(\cdot)$ is base $2$ and $\ln(\cdot)$ is base $e$.

\subsection{Directed Information}
Given a pair of random sequences $X^n$ and $Y^n$,
the \emph{directed information} from $X^n$ to $Y^n$ is defined as
\begin{align}
\DI{X^n}{Y^n} &\triangleq  \sum_{i=1}^n I(X^i ; Y_i|Y^{i-1}) \\
&= H(Y^n) - H(Y^n \| X^n), \label{di-def}
\end{align}
where $H(Y^n \| X^n)$ is the \emph{causally conditional entropy} \cite{Kramer1998}, defined as
\begin{equation}
H(Y^n \| X^n) \triangleq \sum_{i=1}^n H(Y_i|Y^{i-1},X^i).
\end{equation}
Compared to mutual information
\begin{equation}
I(X^n;Y^n)=H(Y^n)-H(Y^n|X^n),
\end{equation}
directed information in~\eqref{di-def} has the causally conditional entropy
in place of the conditional entropy. Thus, unlike mutual information, directed information is not symmetric, i.e., $\DI{Y^n}{X^n}\neq \DI{X^n}{Y^n}$, in general.

The following notation of \emph{causally conditional pmfs} will be used throughout:
\begin{align}
p(x^n\| y^n) &= \prod_{i  = 1}^n p(x_i|x^{i-1},y^i), \\
p(x^n\| y^{n-1}) &= \prod_{i = 1}^n p(x_i|x^{i-1},y^{i-1}).
\end{align}
It can be easily verified that
\begin{equation}
p(x^n,y^n) = p(y^n\| x^n)p(x^n \| y^{n-1})
\end{equation}
and that we have the \emph{conservation laws}:
\begin{align}\label{eqn.conser}
I(X^n;Y^n) & = \DI{X^n}{Y^n} + \DI{Y^{n-1}}{X^n}, \\
I(X^n;Y^n) & = \DI{X^{n-1}}{Y^n} + \DI{Y^{n-1}}{X^n} \nonumber \\
& \quad + \sum_{i = 1}^n I(X_i;Y_i|X^{i-1},Y^{i-1}),
\end{align}
where
\begin{align}
\DI{Y^{n-1}}{X^n} & = \DI{(\emptyset,Y^{n-1})}{X^n} \\
& = H(X^n) - \sum_{i=1}^n H(X_i | X^{i-1}, Y^{i-1})
\end{align}
denotes the \emph{reverse directed information}.
Other interesting properties of directed information can be found in \cite{Kramer1998, Massey--Massey2005, Amblard--Michel2011}.

The \emph{directed information rate} \cite{Kramer1998} between a pair of jointly stationary finite-alphabet processes $\bX$ and $\bY$ is defined as
\begin{equation}
\DIR{\bX}{\bY} \triangleq \lim_{n\to\infty}\frac{1}{n}\DI{X^n}{Y^n}.
\end{equation}
The existence of the limit can be checked \cite{Kramer1998} as
\begin{align}
\DIR{\bX}{\bY}
&= \lim_{n\to\infty}\frac{1}{n}I(X^n\to Y^n)\\
&= \lim_{n\to\infty}\frac{1}{n}\left(H(Y^n)-H(Y^n \| X^n)\right)\\
&= \lim_{n\to\infty}\frac{1}{n}\sum_{i=1}^nH(Y_i|Y^{i-1}) \nonumber\\
& \qquad -\lim_{n\to\infty}\frac{1}{n}\sum_{i=1}^n H(Y_i|Y^{i-1},X^i)\\
&= H(Y_0|Y_{-\infty}^{-1}) - H(Y_0|X_{-\infty}^0,Y_{-\infty}^{-1}),
\end{align}
where the last equality is obtained via the property of Ces\'aro mean
and standard martingale arguments; see \cite[Chs.~4 and~16]{Cover--Thomas2006}. Note that the entropy rate $\rH(\bY)$ of the process $\bY$ is equal to $H(Y_0|Y_{-\infty}^{-1})$.
In a similar vein,
the \emph{causally conditional entropy rate} is defined as
\begin{align}
\rH(\bY \| \bX) &\triangleq \lim_{n\to\infty} \frac{1}{n} H(Y^n \| X^n) \\
&= H(Y_0|X_{-\infty}^0,Y_{-\infty}^{-1}). \label{eqn.causalcond}
\end{align}
Thus,
\begin{equation} \label{eqn.split}
\DIR{\bX}{\bY}=\rH(\bY) - \rH(\bY \| \bX).
\end{equation}
This identity shows that if we estimate $\rH(\bY)$ and $\rH(\bY \| \bX)$ separately and if both estimates converge, we have a convergent estimate of the directed information rate.


\subsection{Universal Probability Assignment}

A probability assignment $Q$ consists of a set of conditional
pmfs $Q(x_i|x^{i-1})$ for every $x^{i-1}\in \cX^{i-1}$ and $i = 1,2,\ldots.$
Note that $Q$ induces a probability measure on a random process $\bX$
and the pmf $Q(x^n) = Q(x_1)Q(x_2|x_1)\cdots Q(x_n|x^{n-1})$ on
$X^n$ for each $n$.

\begin{definition}[Universal probability assignment] \label{def.Q}
Let $\mathscr{P}$ be a class of probability measures.
A probability assignment $Q$ is said to be {\em universal for the class $\mathscr{P}$} if
the normalized relative entropy satisfies
\begin{equation}
\lim_{n\to\infty}\frac{1}{n}D(P(x^n) ||Q(x^n)) = 0
\end{equation}
for every probability measure $P$ in $\mathscr{P}$.
A probability assignment $Q$ is said to be {\em universal} (without a qualifier) if it is universal for the class of stationary probability measures.
\end{definition}

\begin{definition}[Pointwise universal probability assignment] \label{def.PQ}
A probability assignment $Q$ is said to be {\em pointwise universal for $\mathscr{P}$} if
\begin{equation}
\limsup_{n\to\infty}
\left(\frac{1}{n} \log \frac{1}{Q(X^n)} -
\frac{1}{n} \log \frac{1}{P(X^n)}\right) \le 0
\quad\textrm{$P$-a.s.}
\end{equation}
for every $P \in \mathscr{P}$.
A probability assignment $Q$ is said to be {\em pointwise universal} if it is pointwise universal for the class of stationary ergodic probability measures.
\end{definition}

It is well known that there exist universal and pointwise universal probability assignments. Ornstein\cite{Ornstein1978} constructed a pointwise universal probability assignment, which was generalized to Polish spaces by Algoet\cite{Algoet1992}. Morvai, Yakowitz, and Algoet\cite{Morvai--Yakowitz--Algoet1997} used universal source codes to induce a probability assignment and established its universality. Since the quantity $(1/n) \log (1/Q(X^n))$ is generally unbounded, a pointwise universal probability assignment is not necessarily universal. However, if we have a pointwise universal probability assignment, it is easy to construct a probability assignment that is both pointwise universal and universal. Let $Q_1(x^n)$ be a pointwise universal probability assignment and $Q_2(x^n)$ be the i.i.d. uniform distribution, then it is easy to verify that
\begin{equation}
\tilde{Q}(x^n) = a_n Q_2(x^n) + (1-a_n)Q_1(x^n) \label{eqn.both}
\end{equation}
is both universal and pointwise universal provided that $a_n$ decays subexponentially, for example, $a_n = 1/n$. For more discussions on universal probability assignments, see, for example, \cite{MerhavFeder98} and the references therein.


\subsection{Context-Tree Weighting} \label{subsec.ctw}
The sequential probability assignment we use in the implementations of our directed information estimators is the celebrated context-tree weighting (CTW) algorithm
by Willems, Shtarkov, and Tjalken~\cite{Willems--Shtarkov--Tjalkens1995}.
One of the main advantages of the CTW algorithm is that its computational complexity is linear in the block length $n$, and
the algorithm provides the probability assignments $Q$
directly; see \cite{Willems--Shtarkov--Tjalkens1995} and \cite{Willems--Tjalkens1997}.
Note that while the original CTW algorithm was tuned for binary processes, it has been extended for larger alphabets in \cite{Tjalkens--Shtarkov--Willems1993},
an extension that we use in this paper.  In our experiments with simulated data,
we assume that the depth of the context tree is larger than the memory of the source. This assumption can be alleviated by the algorithm introduced by Willems~\cite{Willems1998}, which we will not implement in this paper.

\begin{figure}[htb]
\begin{minipage}[b]{1.0\linewidth}
  \centering
  \psfrag{root}{root}
   \centerline{\includegraphics[height=5.5cm]{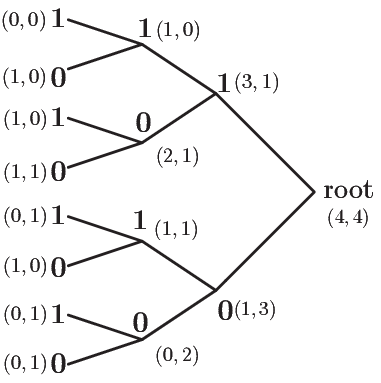}}
\end{minipage}
\centering
\caption{An illustration of the CTW algorithm when $D = 3$ and $(x_{-2}, x_{-1}, x_0, x_1, \ldots, x_8) = 00011010010$. The count starts at $x_1$. For example, there are $3$ zeros and $1$ one with context $1$, represented by count $(3,1)$ at the node of context $1$ in the upper right.}
\label{fig.ctw}
\end{figure}

\begin{figure*}[bh]
\normalsize
\setcounter{equation}{29}
\hrulefill
\begin{align}
\quad P_w^s(X_{n+1} = q|x^n) &= \frac{P_w^s(X_{n+1} = q, x^n)}{P_w^s(x^n)} \\
& = \frac{\frac{1}{2}P_e^s(X_{n+1} = q,x^n)+\frac{1}{2}\prod_{i = 0}^{M-1}P_w^{is}(X_{n+1} = q,x^n)}{P_w^s(x^n)}\\
& = \frac{\frac{1}{2}P_e^s(x^n)P_e^s(X_{n+1}=q|x^n)+\frac{1}{2}P_w^{js}(X_{n+1}=q|x^n)\prod_{i = 0}^{M-1}P_w^{is}(x^n)}{P_w^s(x^n)} \\
& = \frac{\beta^s(x^n)}{1+\beta^s(x^n)}P_e^s(X_{n+1} = q|x^n) + \frac{1}{1+\beta^s(x^n)}P_w^{js}(X_{n+1} = q|x^n),\label{eqn.weighting}
\end{align}
\end{figure*}

\setcounter{equation}{22}

An example of a context tree of input sequence $x_{-2}^8$ with a binary alphabet is shown in Fig.~\ref{fig.ctw}. In general, each node in the tree corresponds to a context, which is a string of symbols preceding the symbol that follows. For concreteness, assume the alphabet is $\{0,1,\ldots,M-1\}$. With a slight abuse of notation, we use $s$ to represent both a node in the context tree and a specific context. At every node $s$, we use a length-$M$ array $(a_{0,s},a_{1,s},\ldots,a_{M-1,s})$ to count the numbers of different values emitted with context $s$ in sequence $x^n$. In Fig.~\ref{fig.ctw}, the counts $(a_{0,s},a_{1,s})$ are marked near each node $s$, and they are simply numbers of zeros and ones emitted from node~$s$.

Take any sequence $z^n$ whose alphabet is $\{0,1,\ldots,M-1\}$. If $z^n$ contains $b_0$ zeros, $b_1$ ones, $b_2$ twos, and so on, the Krichevsky--Trofimov probability estimate of $z^n$\cite{Krichevsky--Trofimov1981}, i.e., $P_e(z^n) = P_e(b_0,b_1,\ldots,b_{M-1})$ can be computed sequentially. We let $P_e(0,0,\ldots,0) = 1$, and for $b_0\geq 0,b_1\geq 0,\ldots,b_{M-1}\geq 0, 0\leq i\leq M-1$, we have
\begin{align}
& P_e(b_0,b_1,\ldots,b_{i-1}, b_i + 1, b_{i+1},\ldots,b_{M-1}) \nonumber \\
& \triangleq \frac{b_i+1/2}{b_0 + b_1 + \ldots + b_{M-1} + M/2}  \nonumber\\
& \quad \times P_e(b_0,b_1,\ldots,b_{i-1},b_i,b_{i+1},\ldots, b_{M-1}). \label{eqn.ktestimate}
\end{align}

We denote the Krichevsky--Trofimov probability estimate of the $M$-array counts at node $s$ of sequence $x^n$ as $P_e^s(x^n)$. The weighted probability $P_w^s$ at node $s$ of sequence $x^n$ in the CTW algorithm is calculated as
\begin{equation}\label{eqn.howtoboundlemma}
P_w^s(x^n) =
\begin{cases}
\frac{1}{2}P_e^s(x^n) + \frac{1}{2}\prod_{i = 0}^{M-1} P_w^{is}(x^n)& 0\leq l(s)<D\\
P_e^s(x^n) & l(s) = D
\end{cases}
\end{equation}
where the node $is$ is the $i^{th}$ child of node $s$ and $l(s)$ is the depth of node $s$. When we build the context tree from sequence $x^n$, we add symbols one by one. In adding symbol $x_t, 1\leq t \leq n$, we have to update the counts $(a_{0,s},a_{1,s},\ldots,a_{M-1,s})$, the estimated probability $P_e^s$, and the weighted probability $P_w^s$ for each context $s$ of $x_t$. The order of updates is from the context of the longest depth (a leaf node) to the root.

Let $\lambda$ denote the root node of the context tree, then $P_w^\lambda(x^n)$ is the universal probability assignment in the CTW algorithm, which will be denoted as $Q(x^n)$ in Section~\ref{sec.estimation}. We compute the sequential probability assignments as
\begin{equation}
Q(x_{n+1}|x^n) = \frac{Q(x^{n+1})}{Q(x^n)} = \frac{P_w^\lambda(x^{n+1})}{P_w^\lambda(x^n)}.
\end{equation}

In \cite[Ch.~5]{Willems--Tjalkens1997}, Willems and Tjalkens introduced a factor $\beta^s(x^n)$ at every node $s$ to simplify the calculation of the sequential probability assignment, which could also help understand how the weighted probabilities are updated when the input sequence $x^n$ grows to $x^{n+1}$.
For each node $s$, we define factor $\beta^s(x^n)$ as
\begin{equation}
\beta^s(x^n) \triangleq \frac{P_e^s(x^n)}{\prod_{i = 0}^{M-1}P_w^{is}(x^n)}.
\label{def.beta}
\end{equation}
Assuming $js$ is a context of $x_{n+1}$, where $0\leq j\leq M-1$. Obviously, any other node $ks,k\neq j$ cannot be a context of $x_{n+1}$. We express $P_w^s(X_{n+1} = q|x^n), q= 0,1,\ldots,M-1$ in (\ref{eqn.weighting}) at the bottom of this page, which
shows that the sequential probability assignment $Q(x_{n+1}|x^n)$ is a weighted summation of the Krichevsky--Trofimov sequential probability assignments, i.e., $P_e^s(X_{n+1} = q|x^n)$ at all nodes of the context tree.

By~(\ref{eqn.ktestimate}), for any node $s$,
\begin{equation}
P^s_e(X_{n+1} = q|x^{n}) \geq \frac{1/2}{n + |\cX|/2} \geq \frac{1}{2n+|\cX|}.\label{eqn.bounded}
\end{equation}
Thus, $Q(x_{n+1}|x^n) = \Omega(1/n)$, or more precisely,
\begin{equation}
Q(x_{n+1}|x^n) \geq \frac{1}{2n+|\cX|}.
\end{equation}

The probability assignment $Q$ in the CTW algorithm is both universal and pointwise universal for the class of stationary irreducible aperiodic finite-alphabet Markov processes. For the proof of universality, see\cite{Willems--Shtarkov--Tjalkens1995}. The pointwise universality is proved
in Lemma~\ref{lemma.as} in Appendix A.

\setcounter{equation}{28}

\section{Four Estimators} \label{sec.estimation}
In this section, we introduce four estimators of the directed information rate $\rI(\bX\to\bY)$ of a pair $(\bX,\bY)$ of jointly stationary ergodic processes with finite alphabets. Let $\mathcal{M}(\cX,\cY)$ be the set of all probability distributions on
$\cX\times\cY$. Define $f$ to be the function that maps a joint
pmf $P(x,y)$ of a random pair $(X,Y)$ to the corresponding
conditional entropy $H(Y|X)$, i.e.,
\begin{equation}
f( P ) \triangleq -\sum_{x,y}
P(x,y)\log P(y|x), \label{eqn.conditional}
\end{equation}
where $P(y|x)$ is the
conditional pmf induced by $P(x,y)$. Take $Q$ as a universal probability assignment, either on processes with $(\cX \times \cY)$-valued components, or with $\cY$-valued components, as will be clear from the context.

\setcounter{equation}{33}

Recall the definition of the directed information from $X^n$ to $Y^n$:
\begin{equation}
I(X^n \to Y^n) = \sum_{i = 1}^n I(X^i;Y_i|Y^{i-1}) = H(Y^n) - H(Y^n \| X^n),
\end{equation}
we define the four estimators as follows:
\begin{align}
\hat{I}_1(X^n \to Y^n)
  &\triangleq \hat{H}_1(Y^n) - \hat{H}_1(Y^n\| X^n), \label{eqn.first} \\
\hat{I}_2(X^n \to Y^n)
  &\triangleq \hat{H}_2(Y^n) -\hat{H}_2(Y^n\| X^n), \label{eqn.second}\\
\hat{I}_3(X^n \to Y^n)
  &\triangleq \frac{1}{n}\sum_{i = 1}^n D\bigl(Q(y_i|X^i,Y^{i-1}) \| Q(y_i|Y^{i-1})\bigr), \label{eqn.third}\\
\hat{I}_4(X^n\to Y^n)
  &\triangleq \frac{1}{n}\sum_{i= 1}^n
D \bigl(Q(x_{i+1},y_{i+1}|X^i,Y^i) \nonumber \\[-1.5em]
& \qquad \qquad \quad\; \| Q(y_{i+1}|Y^i)Q(x_{i+1}|X^i,Y^i)\bigr), \vphantom{\sum_{i=1}^n}\label{eqn.end}
\end{align}
where
\begin{align}
\hat{H}_1(Y^n \| X^n)
  &\triangleq -\frac{1}{n}\log Q(Y^n \| X^n), \label{eqn.aeptype}\\
\hat{H}_2(Y^n \| X^n)
  &\triangleq \frac{1}{n}\sum_{i = 1}^n f(Q(x_{i+1},y_{i+1}|X^i,Y^i)), \\
  \hat{H}_2(Y^n) 
  & \triangleq \frac{1}{n}\sum_{i = 1}^n \sum_{y_{i+1}} Q(y_{i+1}|Y^i)\log \frac{1}{Q(y_{i+1}|Y^i)}, \\
  \hat{H}_1(Y^n)
  & \triangleq \hat{H}_1(Y^n \| \emptyset^n).
\end{align}

Recall that
$Q(y_i | X^i, Y^{i-1})$ denotes the conditional pmf $Q(y_i | x^i,y^{i-1})$ evaluated for the random sequence $(X^i,Y^{i-1})$,
and $Q(Y^n || X^n)$ denotes the causally conditional pmf $Q(y^n||x^n)$ evaluated
for $(X^n,Y^n)$.
Thus, an entropy estimate such as $\hat{H}_1(Y^n \| X^n)$
is a \emph{random variable} (since it is a function of $(X^n, Y^n)$), as opposed to the entropy terms such as $H(Y^n \| X^n)$, which are deterministic and
depend on the \emph{distribution} of $(X^n, Y^n)$.

Note that in~\eqref{eqn.third} and~\eqref{eqn.end} the universal probability
assignments conditioned on different data are calculated separately. For example,
$Q(y_i|Y^{i-1})$ is not computed from $Q(x_i,y_i|X^{i-1},Y^{i-1})$, but from running the universal probability assignment algorithm again on dataset $Y^{i-1}$.
In the case of $Q(Y_i|X^i, Y^{i-1})$, which is inherent in the computation of $Q(Y^n \| X^n)$, the estimate is computed from pmf $Q(x_i, y_i | X^{i-1}, Y^{i-1})$ via
$Q(Y_i|X^i, Y^{i-1}) = Q(X_i, Y_i | X^{i-1}, Y^{i-1})/\sum_{y_i} Q(X_i, y_i | X^{i-1}, Y^{i-1})$.

We can express $\hat{I}_4$ in another form which might be enlightening:
\begin{equation}
\hat{I}_4 = G_n - \hat{H}_2(Y^n \| X^n),
\end{equation}
where $G_n$ is 
\begin{equation}
G_n = \frac{1}{n}\sum_{i= 1}^n\sum_{(x_{i+1},y_{i+1})} Q(x_{i+1},y_{i+1}|X^i,Y^i)\log \frac{1}{Q(y_{i+1}|Y^i)}.
\end{equation}

It is also worthwhile to note that $\hat{I}_4$ involves an average of $x_{i+1}$ in the relative entropy term for each $i$, which makes it analytically different from $\hat{I}_3$.

Here is the big picture of the general ideas behind these estimators. The first estimator $\hat{I}_1$ is calculated through the difference of two terms, each of which takes the form of (\ref{eqn.aeptype}). Since the Shannon--McMillan--Breiman theorem guarantees the asymptotic equipartition property (AEP) of entropy rate\cite{Cover--Thomas2006} as well as directed information rate\cite{Venkataramanan--Pradhan2007}, it is natural to believe that $\hat{I}_1$ would converge to the directed information rate. This is indeed the case, which is proved in Appendix B. The Shannon--McMillan--Breiman type estimators have been widely applied in the literature of information-theoretic measure estimation, for example, relative entropy estimation by Cai, Kulkarni, and Verd\'u \cite{Cai--Kulkarni--Verdu2006}, and erasure entropy estimation by Yu and Verd\'u \cite{Yu--Verdu2006}.

Equation (\ref{eqn.aeptype}) can be rewritten in the Ces\'aro mean form, i.e.,
\begin{equation}
-\frac{1}{n}\log Q(Y^n\|X^n) = \frac{1}{n}\sum_{i = 1}^n \log \frac{1}{Q(Y_i|Y^{i-1},X^i)}, \label{eqn.cesarotype}
\end{equation}
and estimators $\hat{I}_2$ through $\hat{I}_4$ are derived by changing every term in the Ces\'aro mean to other functionals of probability assignments $Q$. For concreteness, estimator $\hat{I}_2$ uses conditional entropy as the functional, and estimators $\hat{I}_3$ and $\hat{I}_4$ use relative entropy.

One disadvantage of $\hat{I}_1$ is that it has a nonzero probability of being very large, since it only averages over logarithms of estimated conditional probabilities, while the directed information rate that it estimates is always bounded by $\log |\mathcal{Y}|$.

The estimator $\hat{I}_2$ is the universal directed information estimator introduced in \cite{Zhao--Kim--Permuter--Weissman2010}. Thanks to the use of information-theoretic functionals to ``smooth'' the estimate, the absolute value of $\hat{I}_2(X^n \to Y^n)$ is upper bounded by $\log |\mathcal{Y}|$ on any realization, a clear advantage over $\hat{I}_1$.

The common disadvantage of $\hat{I}_1$ and $\hat{I}_2$ is that they are computed by subtraction of two nonnegative quantities. When there is insufficient data, or the stationarity assumption is violated, $\hat{I}_1$ and $\hat{I}_2$ may generate negative outputs, which is clearly undesirable. In order to overcome this, $\hat{I}_3$ and $\hat{I}_4$ are introduced, which take the form of a (random) relative entropy and are always nonnegative. Section~\ref{subsec.stock} gives an example where $\hat{I}_1$ and $\hat{I}_2$ give negative estimates, which might be caused by the fact that the underlying process (stock market) is not stationary, at least in a short term.


\section{Performance Guarantees}\label{section.gua}

In this section, we establish the consistency of the proposed estimators, mainly in the almost sure and $L_1$ senses. Under some mild conditions, we derive near-optimal rates of convergence in the minimax sense. The proofs of the stated results are given in the Appendices.

\begin{theorem}\label{thm.main2}
Let $Q$ be a universal probability assignment and $(\bX, \bY)$ be a pair of jointly stationary ergodic
finite-alphabet processes. Then
\begin{equation} \label{eq.i1conv}
\lim_{n\to\infty}\hat{I}_1(X^n \to Y^n)=\rI(\bX\to\bY) \quad\textrm{in $L_1$}.
\end{equation}
Furthermore, if $Q$ is also a pointwise universal probability assignment, then the limit in \eqref{eq.i1conv} holds almost surely as well.
\end{theorem}

The proof of Theorem~\ref{thm.main2} is in Appendix \ref{app.thm.main2}. If $(\bX,\bY)$ is a stationary irreducible aperiodic finite-alphabet Markov process,
we can say more about the performance of $\hat{I}_1$ using the probability assignment in the CTW algorithm.

\begin{proposition}\label{cor.h2}
Let $Q$ be the CTW probability assignment and let $(\bX, \bY)$ be a jointly stationary irreducible aperiodic finite-alphabet Markov process whose order is bounded by the prescribed tree depth in the CTW algorithm, and let $\bY$ be a stationary irreducible aperiodic finite-alphabet Markov process with the same order as $(\bX,\bY)$. Then there exists a constant $C_1$ such that
\begin{equation}
\mathbb{E}\left |\hat{I}_1(X^n \to Y^n) - \rI(\bX\to\bY)\right |\leq C_1n^{-1/2}\log n,
\end{equation}
and $\forall\epsilon>0$, $P$-a.s.
\begin{equation}
\left |\hat{I}_1(X^n \to Y^n) - \rI(\bX\to\bY)\right | = o(n^{-1/2}(\log n)^{5/2 + \epsilon}).
\end{equation}
\end{proposition}

The proof of Proposition~\ref{cor.h2} is in Appendix~\ref{app.cor.h2}.

We can establish similar consistency results for the second estimator $\hat{I}_2$ in~\eqref{eqn.second}.

\begin{theorem} \label{thm.main}
Let $Q$ be a universal probability assignment, and finite-alphabet process $(\bX, \bY)$ be jointly stationary ergodic. Then
\begin{equation}
\lim_{n\to\infty}\hat{I}_2(X^n \to Y^n)=\rI(\bX\to\bY)\mbox{ in $L_1$}.
\end{equation}
\end{theorem}

The proof of Theorem~\ref{thm.main} is in Appendix~\ref{app.thm.main}. As was the case for $\hat{I}_1$, if the process $(\mathbf{X,Y})$ is a jointly stationary irreducible aperiodic finite-alphabet Markov process, we can say more about the performance of $\hat{I}_2$ using the CTW algorithm as follows:

\begin{proposition}\label{cor.hs}
Let $Q$ be the probability assignment in the CTW algorithm. If $(\mathbf{X,Y})$ is a jointly stationary irreducible aperiodic finite-alphabet Markov process whose order does not exceed the prescribed tree depth in the CTW algorithm, and $\bY$ is also a stationary irreducible aperiodic finite-alphabet Markov process with the same order as $(\bX,\bY)$, then
\begin{equation}
\lim_{n\to\infty}\hat{I}_2(X^n \to Y^n)=\rI(\bX\to\bY) \quad\textrm{$P$-a.s. and in $L_1$},
\end{equation}
and there exists a constant $C_2$ such that
\begin{equation}
\mathbb{E}\left |\hat{I}_2(X^n \to Y^n) - \rI(\bX\to\bY)\right |\leq C_2n^{-1/2}(\log n)^{3/2}.
\end{equation}
\end{proposition}

The proof of Proposition~\ref{cor.hs} is in Appendix~\ref{app.cor.hs}.

We also investigate the minimax lower bound of estimating directed information rate, and show the rates of convergence for the first two estimators are optimal within a logarithmic factor. Note that entropy rate is a special case of directed information rate if we take process $\bY = \bX$, so the minimax lower bound also applies in the universal entropy estimation situation. Actually in the proof of proposition~\ref{thm.minimax}, we indeed reduce the general problem to entropy estimation problem to show the minimax lower bound.

\begin{proposition}\label{thm.minimax}
Let $\mathscr{P}(\bX,\bY)$ be any class of processes
that includes the class of i.i.d.\@ processes.
Then, there exists a positive constant $C_3$ such that
\begin{equation}
\inf\limits_{\hat{I}}\sup\limits_{\mathscr{P}(\bX,\bY)}\mathbb{E}|\hat{I} - \DIR{\bX}{\bY}|\geq C_3n^{-1/2},
\end{equation}
where the infimum is over all estimators $\hat{I}$ of the directed information rate based on $(X^n, Y^n)$.
\end{proposition}

The proof of Proposition~\ref{thm.minimax} is in Appendix~\ref{app.thm.minimax}. Evidently, convergence rates better than $O(n^{-1/2})$ is not attainable even with respect to the class of i.i.d. sources and thus, a fortiori, in our setting of a much larger uncertainty set.

For the third and fourth estimators, we establish the following consistency results using the CTW algorithm.

\begin{theorem}\label{thm.hybrid}
Let $Q$ be the probability assignment in the CTW algorithm. If $(\bX, \bY)$ is a jointly stationary irreducible aperiodic finite-alphabet Markov process whose order does not exceed the prescribed tree depth in the CTW algorithm, and $\bY$ is also a stationary irreducible aperiodic finite-alphabet Markov process with the same order as $(\bX,\bY)$, then
\begin{equation}
\lim\limits_{n\to \infty}\hat{I}_3(X^n \to Y^n) = \DIR{\bX}{\bY}\quad\textrm{$P$-a.s. and in $L_1$.}
\end{equation}
\end{theorem}

\begin{theorem}\label{thm.diver}
Let $Q$ be the probability assignment in the CTW algorithm. If $(\bX, \bY)$ is a jointly stationary irreducible aperiodic finite-alphabet Markov process whose order does not exceed the prescribed tree depth in the CTW algorithm, and $\bY$ is also a stationary irreducible aperiodic finite-alphabet Markov process with the same order as $(\bX,\bY)$, then
\begin{equation}
\lim\limits_{n\to \infty}\hat{I}_4(X^n \to Y^n) = \DIR{\bX}{\bY}
\quad\textrm{$P$-a.s. and in $L_1$.}
\end{equation}
\end{theorem}

The proofs of Theorem~\ref{thm.hybrid} and Theorem~\ref{thm.diver} are in Appendices~\ref{app.thm.hybrid} and~\ref{app.thm.diver}.

\begin{remark}
The properties of the CTW probability assignment we use in the proofs of Theorem~\ref{thm.hybrid} and Theorem~\ref{thm.diver} are not only universality and pointwise universality, but also lower boundedness (recall Section II-C).
\end{remark}

\begin{remark}
Note that the assumption that $(\bX, \bY)$ is a jointly stationary irreducible aperiodic finite-alphabet Markov process does not imply $\bY$ also has these properties. Suppose that $\bX$ is a Markov process of order $m$, $\bY$ is a hidden Markov process whose internal process is $\bX$, then it is obvious that joint process $(\bX,\bY)$ is Markov with order $m$, but $\bY$ is not a Markov process. In applications, it is sensible to assume that a process $\mathbf{Z}$ can be approximated by Markov processes better and better as the Markov order increases, i.e., there exists constants $C'>0, 0\leq \rho<1$, such that
\begin{equation}
 0\leq H(Z_0|Z^{-1}_{-k}) - \rH (\mathbf{Z}) \leq \frac{C'}{\ln(2)} \rho^k. \label{eqn.markovapproximation}
\end{equation}
It deserves mentioning that the exponentially fast convergence in (\ref{eqn.markovapproximation}) can be satisfied under mild conditions. For example, as shown in Birch\cite{Birch1962}, let $\mathbf{G}$ be a Markov process with strictly positive transition probabilities, and $Z_n = \psi(G_n)$, then (\ref{eqn.markovapproximation}) holds. For more on this ``exponential forgetting'' property, please refer to Gland and Mevel\cite{Gland--Mevel2000} and Hochwald and Jelenkovi\'c\cite{Hochwald--Jelenkovic1999}.
\end{remark}

The properties established for the proposed estimators are summarized in Table~\ref{table:nonlin}.

\begin{table}[ht]
\caption{Properties of the Proposed Estimators}
\centering
\begin{tabular}{|c | c | c |} 
\hline
& Support & Rates of convergence \\ [0.5ex] %
\hline
$\hat{I}_1$ & $(-\infty,\infty)$  & $O(n^{-1/2}\log n)$\\ [1ex] 
\hline
$\hat{I}_2$ & $[-\log |\mathcal{Y}|,\log|\mathcal{Y}|]$ & $O(n^{-1/2}(\log n)^{3/2})$\\ [1ex] 
\hline 
$\hat{I}_3$ & $[0,\infty)$ & --\\ [1ex] 
\hline
$\hat{I}_4$ & $[0,\infty)$ & --\\ [1ex] 
\hline
\end{tabular}
\label{table:nonlin} 
\end{table}

\section{Algorithms and Numerical Examples}
In this section, we use the context-tree weighting (CTW) algorithm as the universal probability assignment to describe the corresponding directed information estimation algorithms and perform experiments on simulated as well as real data. The CTW algorithm \cite{Willems--Shtarkov--Tjalkens1995} has a linear computational complexity in the block length $n$, and it provides the probability assignment $Q$ directly. A brief introduction on how the CTW works can be found in Section \ref{subsec.ctw}.

For simplicity and concreteness, we explicitly describe the algorithm for computing $\hat{I}_2$. The
algorithms for the other estimators are identical, except for the update rule, which is given, respectively, by (\ref{eqn.first}) to (\ref{eqn.end}).

\begin{algorithm}[h]
\caption{Universal estimator $\hat{I}_2$ based on the CTW algorithm}\label{algorithm}
\begin{algorithmic}[h]
\State Fix block length $n$ and context tree depth $D$.

\State $\hat{I}_2 \gets 0$ \For{$i \gets 1,\, n$}

\State $z_i = (x_i,y_i)$  \Comment{Make a super symbol with alphabet size $|\cX||\cY|$} \EndFor
\For{$i \gets D+1,\, n+1$} \State Gather the context $z^{i-1}_{i-D}$ for the $i$th symbol $z_i$.
 \State Update the context tree for every possible value of $z_i$. The estimated pmf $Q(z_i|Z^{i-1})$
is obtained along the way.

\State Gather the context $y^{i-1}_{i-D}$ for the $i$th symbol $y_i$.

\State Update the context tree for every possible value of $y_i$. The estimated pmf
$Q(y_i|Y^{i-1})$ is obtained along the way.

 \State Update $\hat{I}_2$ as $\hat{I}_2\gets
\hat{I}_2+\funcf{Q(x_i,y_i|X^{i-1},Y^{i-1})}-\funcf{Q(y_i|Y^{i-1})}$ where
$\funcf{\cdot}$ is defined in (\ref{eqn.conditional}).

\EndFor \State $\hat{I}_2\gets \hat{I}_2/(n-D)$
\end{algorithmic}
\end{algorithm}

We now present the performance of the estimators on synthetic and real data. The synthetic data
is generated using Markov processes that are passed through simple channels such as discrete memory
channels (DMC), or channels with intersymbol interference. We compare the performances of the
estimators to each other, as well as the ground truth, which we are able to analytically compute. We also extend the proposed methods to estimation of directed information with delay, and to estimation of mutual information. Further, we show how
one can use the directed information estimator to detect delay of a channel, and to detect the
``causal influence'' of one sequence on another. Finally, we apply our estimators on real stock market data to
detect the causal influence that exists between the Chinese and the US stock
markets.

\subsection{Stationary Hidden Markov Processes}\label{sec.ex1}
\begin{figure}[h]{
\footnotesize
\psfrag{BSC1}[B][][1.2]{\ \ \ \ BSC($\epsilon$)\ \ \ } \psfrag{Xi}[][][1]{${\bf
X}=\{X_i\}_{i=1}^n \ \ \ \ \ \  $} \psfrag{Yi}[][][1]{$\ \ \ \ \ \  {\bf Y}=\{Y_i\}_{i=1}^n$}
\psfrag{Y}[b][][1]{$\ Y_{i-1}$}

\centerline{\includegraphics[width=2.75in]{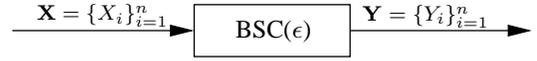}}}

\caption{Section \ref{sec.ex1} setup: $\bX$ is a binary first order Markov process with transition probability $p$, and $\bY$ is the output of a binary symmetric channel with crossover probability $\epsilon$ corresponding to the input $\bX$. }
\label{fig.ex1setup}
\end{figure}

Let $\bX$ be a binary symmetric first order Markov process with transition probability $p$, i.e.
$\PR(X_n\neq X_{n-1}|X_{n-1})=p$. Let $\bY$ be the output
of a binary symmetric channel with
crossover probability $\epsilon$,
corresponding to the input process $\bX$, as depicted in Fig.~\ref{fig.ex1setup}.


\begin{figure}[h]
\footnotesize \psfrag{n}[][][1]{$n$} \psfrag{log Q}[][][1]{$\hat I_1(Y^n \to X^n)$}
\psfrag{entropy fun}[][][1]{$\hat I_2(Y^n \to X^n)$} \psfrag{hybrid}[][][1]{$\hat I_3(Y^n \to
X^n)$} \psfrag{div fun}[][][1]{$\hat I_4(Y^n \to X^n)$}
\centerline{\includegraphics[width=3.5in]{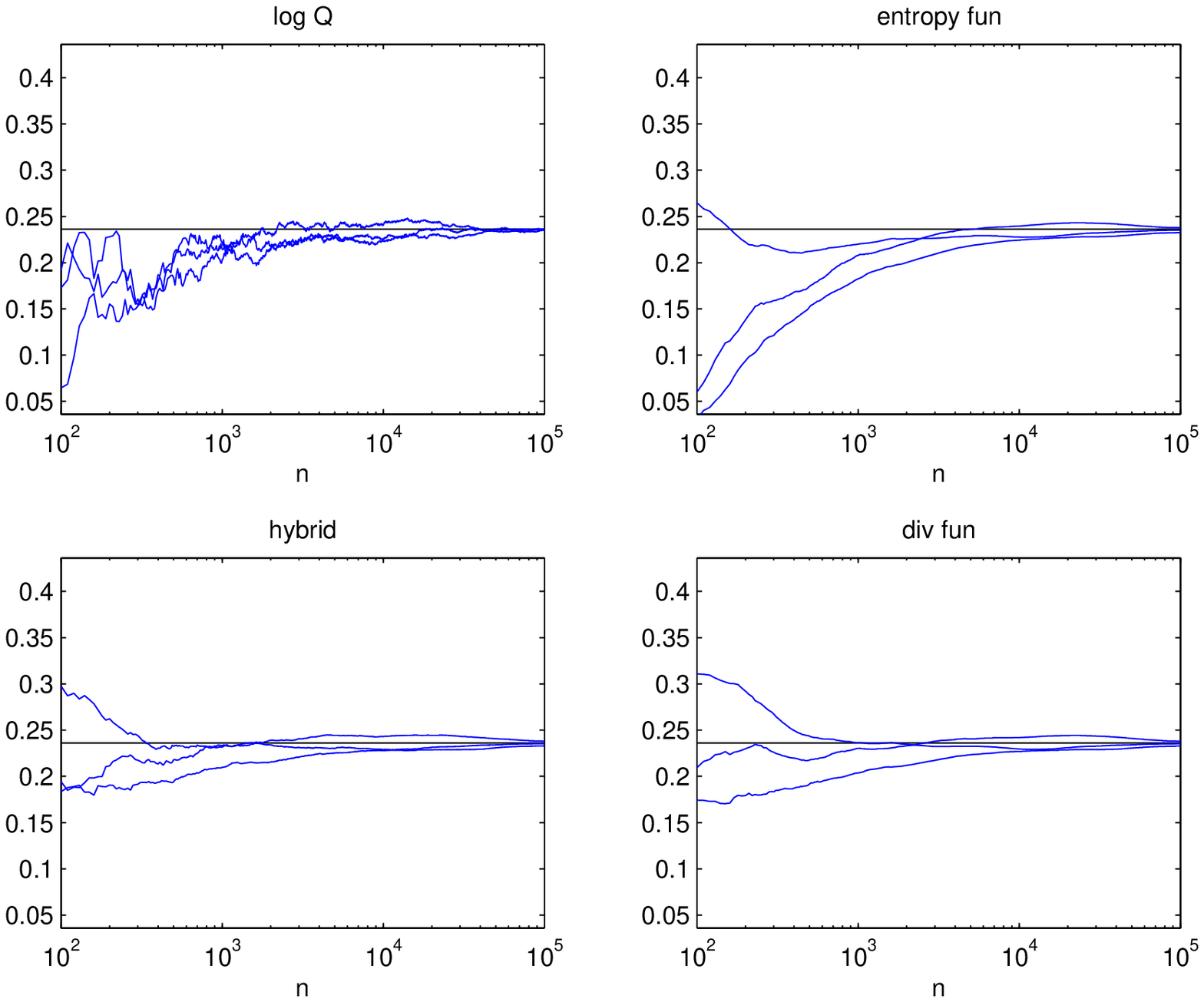}} \caption{Estimation of $\DIR{\bY}{\bX}$:
The straight line is the analytical value.} \label{fig.ex1}
\end{figure}

We use the four algorithms presented to estimate the directed information rate
$\DIR{\bY}{\bX}$ for the case where $p=0.3$ and $\epsilon=0.2$. The depth of the context tree is set to be $3$. The simulation was performed three times. The results are shown in Fig.~\ref{fig.ex1}. As the data length grows, the estimated value approaches the true value for all four algorithms.

The true value can be simply computed analytically as
\begin{align}
I(Y^n\to X^n)&=H(X^n)-H(X^n||Y^n) \\
&= \sum_{i=1}^n H(X_i|X^{i-1})-H(X_i|X^{i-1},Y^i) \\
& = \sum_{i=1}^n H(X_i|X_{i-1})-H(X_i|X_{i-1},Y_i)\label{eqn.iyxa} \\
& =  \sum_{i=1}^n H_b(p)-(p\epsilon+\bar p \bar
\epsilon)H_b\left(\frac{p\epsilon}{p\epsilon+\bar p \bar \epsilon}\right) \nonumber\\
& \quad -(\bar p\epsilon+ p \bar \epsilon)H_b\left(\frac{\bar p\epsilon}{\bar p\epsilon+ p \bar \epsilon}\right),\label{eqn.iyxb}
\end{align}
where (\ref{eqn.iyxa}) follows from the Markov property of the input process and the memorylessness of the
channel and in (\ref{eqn.iyxb}), and $\bar p$ denotes $1-p$.

One can note from Fig.~\ref{fig.ex1} that the sample paths of $\hat{I}_2$ and $\hat{I}_4$ indeed appear to be smoother, as one might expect from that fact that they use
the entropy and relative entropy functionals on the pmf estimate $Q(x_i,y_i|Y^{i-1},X^{i-1})$. The first estimator is apparently the least smooth, since it uses the probability assignments evaluated on the sample path, and is highly sensitive to its idiosyncrasies.

\subsection{Channel Delay Estimation via Shifted Directed Information}\label{subsection.delay}

\begin{figure}[h]
\footnotesize
\psfrag{delay}[][][1]{$\ \  D'$ units delay} \psfrag{x}[][][1]{$\cdots X_{0}, X_{1}, \cdots
\ \ $} \psfrag{y}[][][1]{$\ \ \ \ \ \cdots Y_{0}, Y_{1}, \cdots $}

\psfrag{chan1}[][][1]{A channel} \psfrag{chan2}[][][1]{with memory}
  \centerline{\includegraphics[width=3.0in]{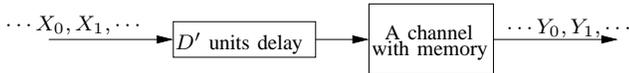}}
\caption{Using the shifted directed information estimation to find the delay $D'$.}
\label{f_delay}
\end{figure}

Assume a setting similar to that in Section~\ref{sec.ex1}, a stationary process that passes through a channel, but now there exists a delay in the entrance of the input to the channel, as depicted
in Fig. \ref{f_delay}.

Our goal is to find the delay $D'$. We use the shifted directed information
$I(Y^{n}_{d+1}\to X^{n-d})$ to estimate $D'$, where $I(Y^{n}_{d+1}\to X^{n-d})$ is defined as
\begin{equation}
I(Y^{n}_{d+1}\to X^{n-d})\triangleq \sum_{i=1}^{n-d} H(X_{i}|X^{i-1})-H(X_{i}|X^{i-1},Y^{d+i}_{d+1}).
\end{equation}

To illustrate the idea, suppose $\bX$ is a binary stationary process, and we define the binary process $\bY$ as follows
\begin{equation}\label{e_channel_memory}
Y_i=X_{i-D'}+X_{i-D'-1}+W_i,
\end{equation}
where $W_i\sim \text{Bernouli}(\epsilon)$ and addition in (\ref{e_channel_memory}) is modulo 2. The goal is to find the delay $D'$ from the
observations of the processes $\bf Y$ and $\bf X$. Note that the mutual information rate $\lim\limits_{n\to \infty}\frac{1}{n} I(Y^n;X^n)$ is not influenced by $D'$. However, the shifted directed information rate $\lim\limits_{n\to \infty} \frac{1}{n-d} I(Y^{n}_{d+1}\to X^{n-d})$ is highly influenced by $D'$. Assuming that there is no feedback, for $d<D'$ we have the Markov chain $Y^{i+d}_{d+1} \to X^{i-1} \to X_i$ due to (\ref{e_channel_memory}), and therefore $I(Y^{n}_{d+1}\to X^{n-d})=0$.
However, for $d\geq D'$, $I(Y^{n}_{d+1}\to X^{n-d})>0.$ For instance, in the channel example
(\ref{e_channel_memory}), if $W_i=0$ with probability 1 then for $d\geq D'$, $I(Y^{n}_{d+1}\to
X^{n-d})=H(X^{n-d})$. Therefore, we can use the shifted directed information $I(Y^{n}_{d+1}\to X^{n-d})$ to
estimate $D'$.

\begin{figure}[h]
\psfrag{d}[][][1]{$d$} \psfrag{value}[][][1]{$\hat I_2(Y^{n}_{d+1} \to X^{n-d})$}
\psfrag{I}[][][1]{}
\centerline{\includegraphics[width=2.6 in]{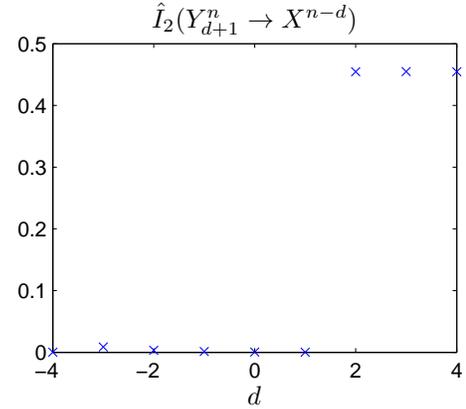}}
\caption{The value of $\hat I_2(Y^{n}_{d+1} \to X^{n-d})$ where  $n=10^6$ for the setting
depicted in Fig. \ref{f_delay} with $D'=2$. When $d<2$, $\hat
I_2(Y^{n}_{d+1} \to X^{n-d})$ is very close to zero and for $d\geq 2$, $\hat I_2(Y^{n}_{d+1} \to X^{n-d})$ is significantly larger than zero. 
} \label{fig.ex2_delay}
\end{figure}

Fig. \ref{fig.ex2_delay} depicts $\hat I_2(Y^{n}_{d+1} \to X^{n-d})$ where $n=10^6$ for the
setting in Fig. \ref{f_delay}, where the input is a binary stationary Markov process of order one
and the channel is given by (\ref{e_channel_memory}). The delay of the channel, $D'$, is equal to 2. We use $\widehat{I}_2$ to estimate the shifted directed information (all algorithms perform similarly for
this case) where the tree depth of the CTW algorithm is set to be 6. The result in Fig. \ref{fig.ex2_delay} shows that for $d<D'$, $\hat
I_2(Y^{n}_{d+1} \to X^{n-d})$ is very close to zero and for $d\geq D'$, $\hat I_2(Y^{n}_{d+1} \to X^{n-d})$ is
significantly larger than zero.

\subsection{Causal Influence Measurement}
There is an extensive literature on detecting and measuring causal influence. See, for example, \cite{Kleinberg--Hripcsak2011} for a recent survey of some
 of the common tools and approaches in biomedical informatics. One particularly celebrated tool, in both the life sciences and economics, for assessing whether and to what extent one time series influences another is the Granger causality test \cite{Granger1969}. The idea is to model $\mathbf{Y}$ first as a univariate autoregressive time series with error correction term $V_i$
\begin{equation}
Y_i = \sum_{j = 1}^p a_jY_{i-j} + V_i, \label{eqn.granger1}
\end{equation}
and then model it again using $\mathbf{X}$ as causal side information:
\begin{equation}
Y_i = \sum_{j = 1}^p [b_j Y_{i-j} + c_j X_{i+1-j}] + \tilde{V_i} \label{eqn.granger2}
\end{equation}
with $\tilde{V_i}$ as the new error correction term. The Granger causality is defined as
\begin{equation}
G_{\bX \to \bY} \triangleq \log \frac{\textrm{var}(V_i)}{\textrm{var}(\tilde{V_i})},
\end{equation}
and the bigger it is, the more inclined the practitioner is to assert that $\textbf{X}$ is causally influencing $\textbf{Y}$. It is a simple exercise to verify that when the process pair is jointly Gauss--Markov with evolution that obeys both (\ref{eqn.granger1}) and (\ref{eqn.granger2}) with $p = \infty$, the Granger causality coincides with the directed information rate (up to a multiplicative constant)\cite{Amblard--Michel2011}.

In this section, we implement our universal estimators of directed information to infer causal influences in more general scenarios, where the Gauss--Markov modeling assumption inherent in Granger causality fails to adequately capture the nature of the data.

\begin{figure}[h!]{
\psfrag{BSC1}[B][][1]{$P(y_i|x^i,y^{i-1})$\ \ } \psfrag{BSC2}[][][1]{$P(x_i|x^{i-1},y^{i-1})$}
\psfrag{Xi}[][][1]{$X_i$} \psfrag{Yi}[][][1]{$Y_i$} \psfrag{Y}[b][][1]{$\ Y_{i-1}$}
\psfrag{Delay}[B][][1]{Delay} \psfrag{t1}[][][1]{\ \ \ forward channel}
\psfrag{t2}[B][][1]{\ \ backward channel}

 \centerline{\includegraphics[width=3in]{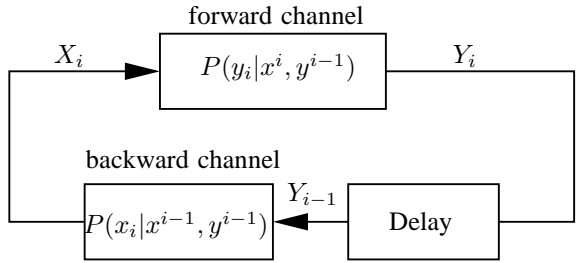}}} \caption{
Modeling any two processes using forward channel $P(y_i|x^i,y^{i-1})$ and backward
channel $P(x_i|x^{i-1},y^{i-1})$.}
\label{f_forward_backward_channel}
 \end{figure}

One philosophical basis for causal analysis is that when we measure causal influence between two processes, $\bX$ and $\bY$, there is an underlying assumption that $X_i$ happens earlier than $Y_i$ for every $(X_i,Y_i)$. Under this assumption, we say two jointly distributed processes $\bX$ and $\bY$ induce a forward channel $P(y_i|x^i,y^{i-1})$ and a backward channel $P(x_i|x^{i-1},y^{i-1})$, as depicted in Fig.~\ref{f_forward_backward_channel}, where $\bX$ is the input process. In this section we
present the use of directed information, reverse directed information, and mutual information to
measure the causal influence between two processes.

 \begin{figure}[h!]{
\psfrag{BSC1}[B][][1]{BSC($\alpha$)\ \ \ } \psfrag{BSC2}[][][1]{BSC($\beta$)}
\psfrag{Xi}[][][1]{$X_i$} \psfrag{Yi}[][][1]{$Y_i$} \psfrag{Y}[b][][1]{$\ Y_{i-1}$}
\psfrag{Delay}[B][][1]{Delay}
 \centerline{\includegraphics[width=3in]{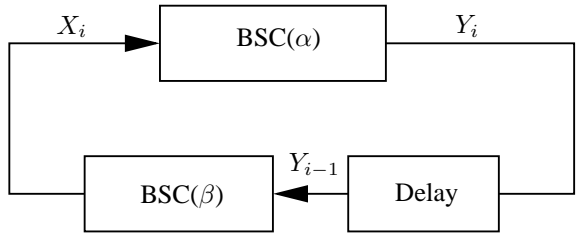}}} \caption{Simulation of a sequence of
random variables $\{X_i,Y_i\}_i\geq 1$ according to the relation shown in the scheme. Namely,
$Y_i$ is the output of a binary symmetric channel with parameter $\alpha$ and input $X_i$ and
$X_i$  is the output of a binary symmetric channel with parameter $\beta$ and input $Y_{i-1}$.
The initial random variable $X_1$ is assumed to be distributed Bernoulli($\frac{1}{2})$.}
\label{figure.haim}
 \end{figure}

\begin{definition}[Existence of a channel]
We say that the forward channel does not exist if $P(y_i|x^i,y^{i-1})=P(y_i|y^{i-1})$ for
$i\geq 1$ and similarly the backward channel does not exist if
$P(x_i|x^{i-1},y^{i-1})=P(x_i|x^{i-1})$ for $i\geq 1$.
\end{definition}

We interpret the existence of the forward link as that the sequence $\bY$ is
``influenced" or ``caused" by the process $\bX$. Similarly, the existence of the
backward link is interpreted as that $\bX$ is ``influenced" or ``caused" by the sequence
$\bY$. We would like to answer the following two questions:
\begin{enumerate}
\item  Does the forward channel exist?
\item Does the backward channel exist?
\end{enumerate}

Directed information can naturally answer these questions. It is straightforward to note from the
definition of directed information that the forward link exists if and only if $I(X^n\to
Y^n)>0$ and the backward link exists if and only if $I(Y^{n-1}\to X^n)>0.$ More generally, the directed information $I(X^n\to Y^n)$ quantifies how much $\bX$
influences $\bY$, while the directed information in the reverse direction $I(Y^{n-1}\to X^n)$ quantifies how much $\bY$ influences $\bX$. The mutual
information, which is the sum of those two directed informations, (see (\ref{eqn.conser})), quantifies the mutual influence of the two sequences. Therefore, using the
directed information measures, it is natural to adopt terminology as follows:
\begin{itemize}
\item Case A: $I(X^n\to Y^n)\gg I(Y^{n-1}\to X^n) $,  we say that  $\bX$ causes $\bY$.
\item Case B: $I(X^n\to Y^n)\ll I(Y^{n-1}\to X^n) $,  we say that  $\bY$ causes $\bX$.
\item Case C: $I(X^n\to Y^n)\simeq I(Y^{n-1}\to X^n) \gg 0$, we say that the processes are mutually causing
each other.
\item Case D: $I(X^n; Y^n)=0$,  we say that the processes are independent of
each other.
\end{itemize}

To illustrate this idea, consider processes $\bX$ and
$\bY$ generated by the system that is depicted in Fig.~\ref{figure.haim},
where the forward channel is a BSC($\alpha$) and the backward channel is a BSC($\beta$) where
$0\leq \alpha \leq \frac{1}{2}$ and $0\leq \beta \leq \frac{1}{2}.$ Intuitively, if $\alpha$ is
much less than $\beta$, then the process $\bX$ is influencing
$\bY$, and if $\alpha$ is much larger than $\beta$, the process
$\bY$ is influencing $\bX$. If $\alpha$ and $\beta$ have
similar values then the processes mutually influence each other, and
finally if they are both equal to $\frac{1}{2}$, then the processes are independent of each
other. Note that the information-theoretic measures can be analytically calculated as in
(\ref{e_analyticalI1})-(\ref{e_analyticalI3}), and indeed if $I(X^n\to Y^n)>I(Y^{n-1}\to X^n)$,
then $\alpha<\beta$ and vice versa. Hence the intuition regarding which process influences the other is consistent with cases A through D presented above.
\begin{equation}\label{e_analyticalI1}
\frac{1}{n}I(X^n\to Y^n) = H_b(\alpha \overline \beta+ \overline \alpha  \beta)-H_b(\alpha)
\end{equation}
where the terms $\overline \alpha$ and $\overline \beta$ denote $1-\alpha$ and $1-\beta$
respectively. Similarly, we have
\begin{equation}
\frac{1}{n}I(Y^{n-1}\to X^n) =H_b(\alpha \overline \beta+ \overline \alpha  \beta)-H_b(\beta)
\end{equation}
and
\begin{equation}\label{e_analyticalI3}
\frac{1}{n}I(Y^n;X^n)=2H_b(\alpha \overline \beta+ \overline \alpha  \beta)-H_b(\beta)-H_b(\alpha).
\end{equation}

Since the normalized reverse directed information is nothing but the normalized directed information between another pair of processes, where one
is shifted, the estimators $\hat{I}_1$ to $\hat{I}_4$ can be easily adapted to this situation, and the convergence theorems (Theorem~\ref{thm.main2} through Theorem~\ref{thm.diver}) apply also (with the appropriate translations) to the reverse directed information. Finally, the normalized mutual information can be estimated once we have the normalized directed
information and the normalized reverse directed information simply by summing them.

\begin{figure}[h]
\psfrag{n}[][][1]{$n$} \psfrag{log Q}[][][1]{Alg. 1} \psfrag{entropy fun}[][][1]{Alg. 2}
\psfrag{hybrid}[][][1]{Alg. 3} \psfrag{div fun}[][][1]{Alg. 4}

\psfrag{MI}[][][0.8]{$\ \ \ \ \ \ \ \ \ \ \ \ \;  \leftarrow \frac{I(X^n;Y^n)}{n}$ }
\psfrag{DI}[][][0.8]{$\ \ \ \ \ \ \ \ \ \ \ \  \; \  \leftarrow  \frac{I(X^n\to Y^n)}{n}$}
 \psfrag{invDI}[][][0.8]{$\ \ \ \ \ \ \ \ \ \ \  \ \; \   \leftarrow  \frac{I(Y^{n-1}\to X^n)}{n}$}

\psfrag{MI1}[][][0.7]{} \psfrag{DI1}[][][0.7]{} \psfrag{invDI1}[][][0.7]{}

\hspace{-1.5em}\centerline{\includegraphics[width=2.8 in]{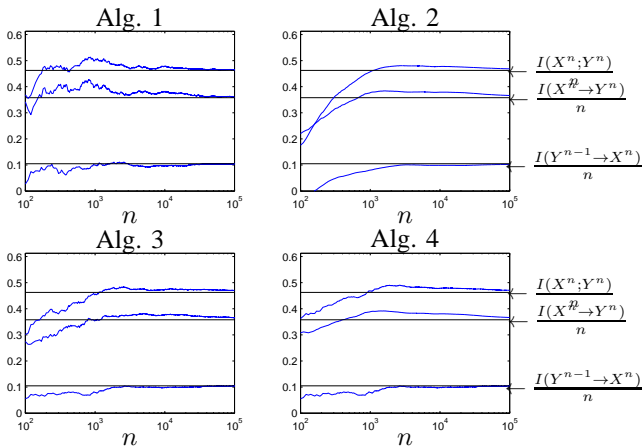}}
\caption{The information-theoretic measures $\frac{1}{n}I(X^n\to Y^n)$, $\frac{1}{n}I(Y^{n-1}\to
X^n)$, and  $\frac{1}{n}I(X^n;Y^n)$ evaluated using the four algorithms. The data was generated
according to the setting in Fig. \ref{figure.haim} where $\alpha=0.1$ and $\beta=0.2$. The
straight black line is the analytical value given by (\ref{e_analyticalI1})-(\ref{e_analyticalI3}) and the blue lines are the estimated values.}
\label{f_all}
\end{figure}

Fig.~\ref{f_all} depicts the estimated and analytical information-theoretic measures
$\frac{1}{n}I(X^n\to Y^n)$, $\frac{1}{n}I(Y^{n-1}\to X^n)$, and  $\frac{1}{n}I(X^n;Y^n)$ for the case $\alpha=0.1$ and $\beta=0.2$. One can note that with just a few hundreds of samples, directed
information and reverse directed information start strongly indicating that $\alpha<\beta$, in other words, $\bX$ influences $\bY$ more than the other way around.


\subsection{Causal Influence in Stock Markets} \label{subsec.stock}
\begin{figure}[h]
\centerline{\includegraphics[width=2.5in]{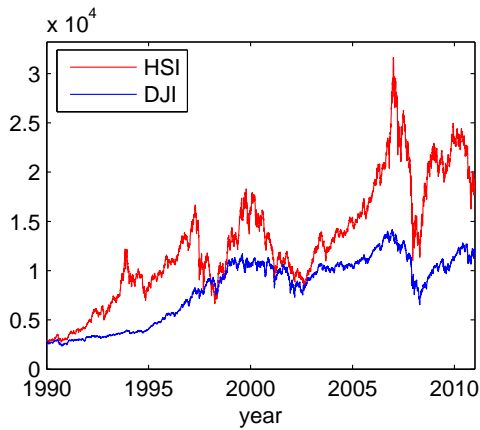}}
\caption{The Hang Seng Index (HSI) and the Dow Jones Industrial Average (DJIA) between 1990 and 2011. The goal is
to determine which index is causally influencing the other.} \label{f_stocks_HSI_DJI}
\end{figure}

Here we use the historic data of the Hang Seng Index (HSI)
and the Dow Jones Index (DJIA) between 1990 and 2011
to compute the directed information rate between these two indexes.
The data of those two indexes
are presented in Fig. \ref{f_stocks_HSI_DJI} on a daily time scale.

There is no time overlap between the stock market in Hong Kong and that in New York, that is, when
the stock market in Hong Kong is open, the stock market in New York is closed, and vice versa. Therefore the
causal influence between the markets is well defined. Since the value of the stock market is
continuous, we discretize it into three values: $-1$, $0$, and $1$. Value $-1$ means that the stock
market went down in one day by more than 0.8\%, value $1$ means that the stock market went up in one
day by more than 0.8\%, and value $0$ means that the absolute change is less than 0.8\%.

\begin{figure}[h]
\footnotesize
  \psfrag{last}[][][1]{Alg. 1} \psfrag{all}[][][1]{Alg. 2}
\psfrag{Hybrid}[][][1]{Alg. 3} \psfrag{DIV FUN}[][][1]{Alg. 4}

\psfrag{MI}[][][0.7]{$\ \ \ \ \ \ \ \ \ \ \ \ \ I(X^n;Y^n)/n$}

\psfrag{DI}[][][0.7]{$\ \ \ \ \ \ \ \ \ \ \ \ \ \ \ \ I(X^n\to Y^n)/n$}

\psfrag{INVDI123456789}[][][0.7]{$I(Y^{n-1}\to X^n)/n$}
    \centering
        \includegraphics[width = 3.5 in]{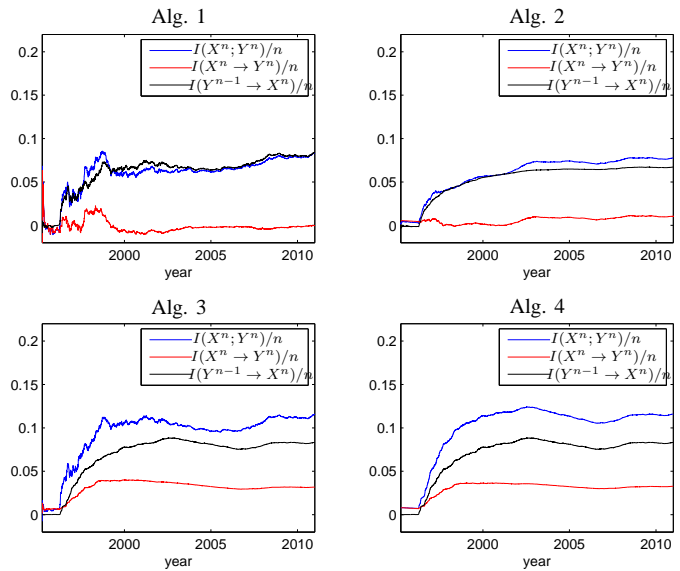}%

    \caption{Estimates of information-theoretic measures between HSI denoted by $\bX$, and DJI denoted by $\bY$. It is clear that the reverse
    directed information is much higher than the directed information, hence it is DJI that causally influences HSI rather than the other way around.}
    \label{f_stock_DI}
\end{figure}

We denote by $X_i$ and $Y_i$ the (quantized ternary valued) change in the HSI and the DJIA in day $i$,
respectively, and estimate the normalized mutual information $\frac{1}{n} I(X^n;Y^n)$, the normalized directed
information $\frac{1}{n} I(X^n \to Y^n)$, and the normalized reverse directed information
$\frac{1}{n}I(Y^{n-1}\to X^n)$, using all four algorithms. Fig.~\ref{f_stock_DI} plots our
estimates of these information-theoretic measures.

Evidently, the reverse directed information is much higher than the directed
information; hence there is a significant causal influence by the DJIA on the HSI,
and a low influence in the reverse direction. In other words, between 1990 and 2011, it was the Chinese market that was influenced by the US market rather than the other way around.

It is also worth noting that estimators $\hat{I}_1$ and $\hat{I}_2$ do generate negative outputs as shown in Fig.~\ref{f_stock_DI}. It may be caused by various reasons, such as data insufficiency and non-stationarity of process $(\bX,\bY)$. In such cases of insufficient data, we would prefer estimators $\hat{I}_3$ and $\hat{I}_4$, since they are always nonnegative, which can be sensibly interpreted in practice.

\section{Concluding Remarks}
We have presented four approaches to estimating the directed information rate between a pair of jointly stationary ergodic finite-alphabet processes. Weak and strong consistency results have been established for all four estimators, in precise senses of varying strengths. For two of these estimators we established convergence rates that are optimal to within logarithmic factors. The other two have their own merits, such as nonnegativty on every sample path. Experiments on simulated and real data substantiate the potential of the proposed approaches in practice and the efficacy of directed information estimation as a tool for detecting and quantifying causality and delay.

\section{Acknowledgments}
The authors would like to thank Todd Coleman for helpful discussions on the merits of nonnegative directed information estimators during Haim Permuter's visit at UCSD. They would like to thank the associate editor and anonymous reviewers for their very helpful suggestions that significantly improved the presentation of our results. Jiantao Jiao would like to thank Hyeji Kim for very helpful discussions in the revision stage of the paper.

\appendices

\section{Some Key Lemmas}
Here is the roadmap of the Appendices. In Appendix A we list some key lemmas without proofs, and in Appendix B we prove the main theorems and propositions in Section \ref{section.gua}. Appendix C provides the proofs of the lemmas in Appendix~A.

The first lemma is on the asymptotic equipartition property (AEP) for causally conditional entropy rate. It was proved in \cite{Venkataramanan--Pradhan2007} that the AEP for causally conditional entropy rate holds in the almost sure sense. Here we prove that it holds in the $L_1$ sense as well. We also show convergence rates for jointly stationary irreducible aperiodic Markov processes.

\begin{lemma} \label{lemma.causalaep}
Let $(\bX, \bY)$ be a jointly stationary ergodic finite-alphabet process. Then,
\begin{equation}
\lim_{n\to\infty} -\frac{1}{n} \log P(Y^n \| X^n) = \rH(\bY \| \bX)\quad \textrm{$P$-a.s. and in $L_1$.}
\end{equation}
In addition, if $(\bX, \bY)$ is irreducible aperiodic Markov, then
\begin{equation}
\mathbb{E}\left |-\frac{1}{n} \log P(Y^n \| X^n) - \rH(\bY \| \bX) \right |  = O(n^{-1/2}\log n)
\end{equation}
and for every $\epsilon>0$,
\begin{align}
& -\frac{1}{n} \log P(Y^n \| X^n) - \rH(\bY \| \bX) \nonumber\\
&\qquad = o(n^{-1/2}(\log n)^{5/2+\epsilon})\quad \textrm{$P$-a.s.}
\end{align}
\end{lemma}

The next lemma shows that the conditional probability induced by the CTW algorithm converges
to the true probability of a Markov process if the CTW depth is sufficiently large.

\begin{lemma}\label{lemma.as}
Let $Q$ be the CTW probability assignment and 
let $\bX$ be a stationary irreducible aperiodic finite-alphabet Markov process whose order is bounded by the prescribed tree depth of the CTW algorithm. Then,
\begin{equation}
\lim_{n\to\infty} Q(x_{n+1}|X^{n}) - P(x_{n+1}|X^{n}) = 0\quad \textrm{$P$-a.s.}
\end{equation}
\end{lemma}

\begin{lemma}[\!\!{{\cite[Lemma 1]{Zhao--Kim--Permuter--Weissman2010}}}]\label{lemma.unifconv}
For any $\epsilon>0$, there exists $K_\epsilon>0$ such that for all $P$ and $Q$ in $\mathcal{M}(\mathcal{X}, \mathcal{Y})$:
\begin{equation}
|f(P)-f(Q)|\leq \epsilon + K_\epsilon \norm{P-Q}_1,
\end{equation}
where $\norm{\cdot}_1$ is the $l_1$ norm (viewing $P$ and $Q$ as $|\mathcal{X}||\mathcal{Y}|$-dimensional simplex vectors), and $f$ is defined in (\ref{eqn.conditional}).
\end{lemma}

\begin{lemma}\label{lemma.diff}
Let $P, Q$ be two probability mass functions in $\mathcal{M}(\mathcal{X}, \mathcal{Y})$, denote $\theta = \| P-Q \|_1$. If $\theta<1/2$, we have
\begin{equation}
|f(P)-f(Q)|\leq 2\theta \log \frac{|\mathcal{X}||\mathcal{Y}|}{\theta},
\end{equation}
where $f$ is defined in (\ref{eqn.conditional}).
\end{lemma}

\begin{lemma} \label{lemma.mixing}
Let $\mathbf{X}$ be a stationary irreducible aperiodic finite-alphabet Markov process. For fixed $i\geq 1$, let random variable $V_i(X^i_{i-m})$ be a deterministic function of random vector $X^i_{i-m}$, where $m$ is the Markov order. Let $V_i$ be uniformly bounded by a constant $V$ for any $i$, and $\mathbb{E}V_i = 0,\forall i\geq 1$. Then there exists a constant $C_4$ such that
\begin{equation}
\mathbb{E}\left ( \frac{1}{n}\sum_{i = 1}^n V_i \right )^2\leq C_4V^2 n^{-1}.
\end{equation}
\end{lemma}

\begin{lemma}[Breiman's generalized ergodic theorem \cite{Breiman1957}]\label{lemma.seconv}
Let $\bX$ be a stationary ergodic process. If $\lim\limits_{k\to\infty}g_k(\bX)= g(\bX)\quad \textrm{$P$-a.s.}$ and $\mathbb{E}[\sup\limits_{k} |g_k|]<\infty$, then
\begin{equation}
\lim_{n\to\infty}\frac{1}{n}\sum_{k=1}^n g_k\left(T^k(\bX)\right)=\E g(\bX) \quad \textrm{$P$-a.s.}
\end{equation}
where $T(\cdot)$ is the shift operator which increases the index by 1, and $T^k$ increases the index by k.
\end{lemma}

Here we paraphrase a result from \cite{Tjalkens--Shtarkov--Willems1993} on the redundancy bounds of the CTW probability assignment.

\begin{lemma}[\!\!\cite{Tjalkens--Shtarkov--Willems1993}]\label{lemma.ctwredundancy}

Let $Q$ be the CTW probability assignment and let $\bX$ be a stationary finite-alphabet Markov process whose order is bounded by the prescribed tree depth of the CTW algorithm. Then there exist constants $C_5,C_6$ such that the pointwise redundancy is bounded as
\begin{equation}
\max\limits_{x^n}\left(\log \frac{1}{Q(x^n)} - \log \frac{1}{P(x^n)} \right)\leq C_5 \log n + C_6 \label{eqn.ctwredundancybounds}
\end{equation}
where $C_5>0, C_6$ depend on nothing but the parameters specifying the process $\mathbf{X}$. In particular, taking expectation over the inequality with respect to $P$, the redundancy is bounded as
\begin{equation}
D(P(x^n)\|Q(x^n))\leq C_5 \log n + C_6. \label{eqn.ctwredundancybounds2}
\end{equation}
\end{lemma}

\begin{remark}
The constants $C_5, C_6$ can be specified once the parameters of process $\mathbf{X}$ are given. For example, see \cite{Tjalkens--Shtarkov--Willems1993}, where
\begin{align}
C_5 & = \frac{(\gamma-1)|\mathcal{S}|}{2}, \\
C_6 & = \frac{(\gamma-1)|\mathcal{S}|}{2}\log\frac{1}{|\mathcal{S}|}+|\mathcal{S}|\left( \frac{\gamma}{\gamma-1} + \log \gamma \right )-\frac{1}{\gamma-1}.
\end{align}
Here $\gamma$ is the size of alphabet, in this case $\gamma = |\mathcal{X}|$. $|\mathcal{S}|$ is the number of states in the Markov process, given Markov order $m$, $|\mathcal{S}|\leq |\mathcal{X}|^{m}$.
\end{remark}

\section{Proofs of Theorems and Propositions}
For brevity, in the sequel we denote $\hat{H}_1(Y^n \| X^n)$ by $\hat{H}_1$, $\hat{H}_2(Y^n \| X^n)$ by $\hat{H}_2$, $\hat{I}_i(X^n \to Y^n)$ by $\hat{I}_i, i = 1,2,3,4$.

\subsection{Proof of Theorem~\ref{thm.main2}} \label{app.thm.main2}
Briefly speaking, we need to show estimator $\hat{I}_1$ converges to the corresponding directed information rate $\DIR{\bX}{\bY}$ for any jointly stationary ergodic process $(\mathbf{X,Y})$. Since $\hat{I}_1$ is defined in (\ref{eqn.first}) as $\hat{H}_1(Y^n)-\hat{H}_1(Y^n\|X^n)$, if we can show the corresponding convergence properties of $\hat{H}_1(Y^n \| X^n)$, then we have the desired convergence properties of $\hat{I}_1$ since $\hat{H}_1(Y^n) = \hat{H}_1(Y^n \| \emptyset)$.

Given $Q$ is a universal probability assignment, first we show $\hat{I}_1$ converges in $L_1$. Then we show given $Q$ is a pointwise universal probability assignment, $\hat{I}_1$ also converges almost surely.

\subsubsection{$L_1$ convergence}
We decompose
\begin{equation}
\hat{H}_1 - \rH(\bY \| \bX) = C_n + D_n, \label{eqn.firstdecomposition}
\end{equation}
where
\begin{equation}
C_n = \hat{H}_1 + \frac{1}{n}\log P(Y^n\| X^n) \label{eqn.cn}
\end{equation}
\begin{equation}
D_n = -\frac{1}{n}\log P(Y^n\| X^n)- \rH(\bY \| \bX). \label{eqn.dn}
\end{equation}
According to Lemma~\ref{lemma.causalaep} shown in Appendix A, we know $D_n$ converges to zero in $L_1$. Now we deal with $C_n$. Pinsker\cite{Pinsker1964} proved the existence of a universal constant $\Gamma>0$ such that
\begin{equation}
D(P\| Q) \leq \mathbb{E}_{P}\left\{\left|\log(\frac{dP}{dQ}) \right|\right\}\leq D(P\| Q) + \Gamma \sqrt{D(P\| Q)}\label{eqn.pinsker},
\end{equation}
Barron\cite{Barron1986} simplified Pinsker's argument and proved that the constant $\Gamma = \sqrt{2}$ is best possible when natural logarithms are used in the definition of $D(P\| Q)$. Here we follow Barron's arguments to bound $\mathbb{E}|C_n|$ with $C_n$ defined in (\ref{eqn.cn}).

Denote the set $\{(x^n,y^n):P(y^n\| x^n)\leq Q(y^n\| x^n)\}$ as $\mathcal{B}_n$, we have
\begin{align}
\mathbb{E}\left|C_n\right| & = \sum_{(x^n,y^n)\in(\mathcal{X\times Y})^n\backslash \mathcal{B}_n}P(x^n,y^n)\frac{1}{n}\log \frac{P(y^n\| x^n)}{Q(y^n\| x^n)} \nonumber \\
&\quad  +\sum_{(x^n,y^n)\in \mathcal{B}_n}P(x^n,y^n)\frac{1}{n}\log \frac{Q(y^n\| x^n)}{P(y^n\| x^n)}\\
& = \mathbb{E}\left[\frac{1}{n}\log \frac{P(Y^n\| X^n)}{Q(Y^n\| X^n)} \right] \nonumber\\
&\quad  + 2\sum_{(x^n,y^n)\in \mathcal{B}_n}P(x^n,y^n)\frac{1}{n}\log \frac{Q(y^n\| x^n)}{P(y^n\| x^n)}
\end{align}

Define $C_{n1} \triangleq \mathbb{E}\left[\frac{1}{n}\log \frac{P(Y^n\| X^n)}{Q(Y^n\| X^n)} \right]$, $C_{n2} \triangleq \sum_{(x^n,y^n)\in \mathcal{B}_n}P(x^n,y^n)\frac{1}{n}\log \frac{Q(y^n\| x^n)}{P(y^n\| x^n)}$, we bound
\begin{align}
C_{n1} & = \frac{1}{n}\sum_{i = 1}^n \mathbb{E}\left [\log\frac{P(Y_i|X^i,Y^{i-1})}{Q(Y_i|X^i,Y^{i-1})} \right]   \\
& = \frac{1}{n}\sum_{i = 1}^n \mathbb{E}\left [\mathbb{E}\left [\log\frac{P(Y_i|X^i,Y^{i-1})}{Q(Y_i|X^i,Y^{i-1})} \Bigg | X^{i-1},Y^{i-1} \right] \right] \\
& \leq \frac{1}{n}\sum_{i = 1}^n \mathbb{E}\left [\mathbb{E}\left [\log\frac{P(Y_i,X_i|X^{i-1},Y^{i-1})}{Q(Y_i,X_i|X^{i-1},Y^{i-1})} \Bigg | X^{i-1},Y^{i-1} \right] \right] \\
\hspace{-0.4cm}& = \frac{1}{n}D(P(x^n,y^n)\| Q(x^n,y^n)). \label{eqn.cn1}
\end{align}
Then, $\forall i$, define $\mathcal{C}_i \triangleq \mathcal{C}_i(x^i,y^{i-1})=\{y_i:P(y_i|x^i,y^{i-1})\leq Q(y_i|x^i,y^{i-1})\}$. We bound $C_{n2}$ from (\ref{eqn.cn2first}) to (\ref{eqn.float1}), where
\begin{itemize}
\item (\ref{eqn.cn2a}) follows by the log-sum inequality~\cite[Theorem 2.7.1]{Cover--Thomas2006},
\item (\ref{eqn.cn2b}) follows since $\forall x>-1, \log(1+x)\leq x/\ln(2)$,
\item (\ref{eqn.cn2c}) follows since $|x|\geq x$,
\item (\ref{eqn.cn2d}) follows by Scheff\'e's theorem~\cite[Lemma 2.1]{Tsybakov2008},
\item (\ref{eqn.cn2e}) follows by Pinsker's inequality~\cite[Lemma 2.5]{Tsybakov2008},
\item (\ref{eqn.cn2f}) follows by the concavity of $\sqrt{\cdot}$,
\item (\ref{eqn.cn2g}) follows by data-processing inequality~\cite[Theorem 2.8.1]{Cover--Thomas2006},
\item (\ref{eqn.float1}) follows by the chain rule for relative entropy, the concavity of $\sqrt{\cdot}$, and data-processing inequality.
\end{itemize}

\begin{figure*}[b!]
\setcounter{equation}{111}
\hrulefill
\begin{align}
C_{n2} & \leq \frac{1}{n} \sum_{i = 1}^n \sum_{(x^i,y^{i-1})} P(x^i,y^{i-1})\sum_{y_i\in \mathcal{C}_i}P(y_i|x^i,y^{i-1})\log \frac{Q(y_i|x^i,y^{i-1})}{P(y_i|x^i,y^{i-1})} \label{eqn.cn2first} \\
& \leq \frac{1}{n} \sum_{i = 1}^n \sum_{(x^i,y^{i-1})}P(x^i,y^{i-1})P(Y_i\in \mathcal{C}_i|x^i,y^{i-1})\log\frac{Q(Y_i\in \mathcal{C}_i|x^i,y^{i-1})}{P(Y_i\in \mathcal{C}_i|x^i,y^{i-1})}\label{eqn.cn2a}\\
& \leq \frac{1}{n} \sum_{i = 1}^n \sum_{(x^i,y^{i-1})}P(x^i,y^{i-1})\frac{1}{\ln(2)}(Q(Y_i\in \mathcal{C}_i|x^i,y^{i-1})-P(Y_i\in \mathcal{C}_i|x^i,y^{i-1}))\label{eqn.cn2b}\\
& \leq  \frac{1}{n} \sum_{i = 1}^n \sum_{(x^i,y^{i-1})}P(x^i,y^{i-1})\frac{1}{\ln(2)}|Q(Y_i\in \mathcal{C}_i|x^i,y^{i-1})-P(Y_i\in \mathcal{C}_i|x^i,y^{i-1})|\label{eqn.cn2c}\\
& \leq  \frac{1}{2n} \sum_{i = 1}^n \sum_{(x^i,y^{i-1})}P(x^i,y^{i-1})\frac{1}{\ln(2)}\sum_{y_i}|P(y_i|x^i,y^{i-1})-Q(y_i|x^i,y^{i-1})|\label{eqn.cn2d}\\
& \leq  \frac{1}{2n} \sum_{i = 1}^n \sum_{(x^i,y^{i-1})}P(x^i,y^{i-1})\sqrt{\frac{2}{\ln(2)}D(P(y_i|x^i,y^{i-1})\| Q(y_i|x^i,y^{i-1})}\label{eqn.cn2e}\\
& \leq  \frac{1}{2n}  \sum_{i = 1}^n \sqrt{\frac{2}{\ln(2)}}\sqrt{\mathbb{E}D(P(y_i|X^i,Y^{i-1})\| Q(y_i|X^i,Y^{i-1})}\label{eqn.cn2f}\\
& \leq  \frac{1}{2n}  \sum_{i = 1}^n \sqrt{\frac{2}{\ln(2)}}\sqrt{\mathbb{E}D(P(y_i,x_{i+1}|X^i,Y^{i-1})\| Q(y_i,x_{i+1}|X^i,Y^{i-1})}\label{eqn.cn2g}\\
& \leq \sqrt{\frac{1}{2\ln(2)}}\sqrt{D(P(x^{n+1},y^{n+1})\| Q(x^{n+1},y^{n+1})/n}\label{eqn.float1},
\end{align}
\end{figure*}

\setcounter{equation}{89}

Combining (\ref{eqn.cn1}) and (\ref{eqn.float1}), we have
\begin{align}
\mathbb{E}\left|C_n\right| & \leq \frac{1}{n}D(P(x^n,y^n)\| Q(x^n,y^n))\nonumber\\
&\quad  + \sqrt{\frac{2}{\ln(2)}}\sqrt{D(P(x^{n+1},y^{n+1})\| Q(x^{n+1},y^{n+1}))/n},\label{bound.l1}
\end{align}
by definition of universal probability assignment, we show $C_n$ converges to zero in $L_1$. Since
\begin{equation}
\mathbb{E}|\hat{H}_1 - \rH(\bY \| \bX)| \leq  \mathbb{E}|C_n| +\mathbb{E}|D_n| \to 0\quad n\to \infty,
\end{equation}
we know $\hat{I}_1$ converges to $\DIR{\bX}{\bY}$ in $L_1$.

\subsubsection{Almost sure convergence}
Consider the probability of the following event
\begin{equation}
\mathcal{A}_{n,\epsilon} = \{(x^n,y^n):\hat{H}_1 \leq -\frac{1}{n}\log P(y^n\| x^n)-\epsilon\}, \label{eqn.typicallower}
\end{equation}
we have
\begin{align}
\mathbb{P}(\mathcal{A}_{n,\epsilon}) & = \sum_{(x^n,y^n)\in \mathcal{A}_{n,\epsilon}}P(x^n,y^n)  \\
& = \sum_{(x^n,y^n)\in \mathcal{A}_{n,\epsilon}} P(y^n\| x^n)P(x^n\| y^{n-1})\\
& \leq \sum_{(x^n,y^n)\in \mathcal{A}_{n,\epsilon}} Q(y^n\| x^n)2^{-n\epsilon}P(x^n\| y^{n-1}) \\
& = 2^{-n\epsilon}\sum_{(x^n,y^n)\in \mathcal{A}_{n,\epsilon}}Q(y^n\| x^n)P(x^n\| y^{n-1}) \\
& \leq 2^{-n\epsilon},
\end{align}
where the first inequality is because of the definition of even $\mathcal{A}_{n,\epsilon}$, and the last step follows from the fact that for any two conditional distributions of the form $Q(y^n\| x^n)$ and $P(x^n\| y^{n-1})$, we have $Q(y^n\| x^n)P(x^n\| y^{n-1}) = \tilde{Q}(x^n,y^n)$ where $\tilde{Q}$ is a joint distribution. As
\begin{equation}
\sum_{n = 1}^{\infty} \mathbb{P}(\mathcal{A}_{n,\epsilon}) <\infty,
\end{equation}
by the Borel-Cantelli Lemma, we have
\begin{equation}
\lim\inf\limits_{n\to \infty}\hat{H}_1 - \left (-\frac{1}{n}\log P(Y^n\| X^n)\right ) \geq 0.\quad \textrm{$P$-a.s.} \label{borel.inf}
\end{equation}
In order to get an inequality with the reverse direction, write $\hat{H}_1 - (-\frac{1}{n}\log P(Y^n\| X^n))$ explicitly as
\begin{align}
&\hat{H}_1 + \frac{1}{n}\log P(Y^n\| X^n) \nonumber\\
&=  \frac{1}{n}\log \frac{P(Y^n\| X^n)}{Q(Y^n\| X^n)}  \\
& =  \frac{1}{n}\log \frac{P(Y^n,X^n)}{Q(Y^n,X^n)}-\frac{1}{n}\log \frac{P(X^n\| Y^{n-1})}{Q(X^n\| Y^{n-1})}, \label{eqn.expansion}
\end{align}
by the definition of pointwise universality (\ref{def.PQ}), we know
\begin{equation}
\lim\sup\limits_{n\to \infty}\frac{1}{n}\log \frac{P(Y^n,X^n)}{Q(Y^n,X^n)} \leq 0,\quad \textrm{$P$-a.s.}
\end{equation}
with a similar argument used for showing (\ref{borel.inf}), we show
\begin{equation}
\lim\sup\limits_{n\to \infty}-\frac{1}{n}\log \frac{P(X^n\| Y^{n-1})}{Q(X^n\| Y^{n-1})}\leq 0,\quad \textrm{$P$-a.s.}
\end{equation}
then we have
\begin{equation}
\lim\sup\limits_{n\to \infty}\hat{H}_1 - \left (-\frac{1}{n}\log P(Y^n\| X^n)\right ) \leq 0.\quad \textrm{$P$-a.s.}\label{eqn.inversedirection}
\end{equation}
Combining (\ref{eqn.inversedirection}) with (\ref{borel.inf}),
\begin{equation}
\lim\limits_{n\to \infty}\hat{H}_1 - \left (-\frac{1}{n}\log P(Y^n\| X^n)\right ) = 0.\quad \textrm{$P$-a.s.}
\end{equation}
By Lemma~\ref{lemma.causalaep} shown in Appendix A,
\begin{equation}
\lim\limits_{n\to \infty}-\frac{1}{n}\log P(Y^n\| X^n)  - \rH(\bY \| \bX) = 0, \quad \textrm{$P$-a.s.}
\end{equation}
which implies the convergence of $\hat{I}_1$ to $\DIR{\bX}{\bY}$ also holds almost surely.

\subsection{Proof of Proposition~\ref{cor.h2}}\label{app.cor.h2}
For similar reasons as shown in the proof of Theorem~\ref{thm.main2}, here it suffices to show the convergence properties of $\hat{H}_1$. For convenience, we restate some arguments shown in the proof of Theorem~\ref{thm.main2}. We decompose $\hat{H}_1 - \rH(\bY \| \bX)$ as
\begin{equation}
\hat{H}_1 - \rH(\bY \| \bX) = C_n + D_n, \label{eqn.firstdecompositionrepeat}
\end{equation}
where
\begin{equation}
C_n = \hat{H}_1 + \frac{1}{n}\log P(Y^n\| X^n) \label{eqn.cnrepeat}
\end{equation}
\begin{equation}
D_n = -\frac{1}{n}\log P(Y^n\| X^n)- \rH(\bY \| \bX), \label{eqn.dnrepeatnew}
\end{equation}
and we restate (\ref{bound.l1})
\begin{align}
\mathbb{E}\left|C_n\right| & \leq \frac{1}{n}D(P(x^n,y^n)\| Q(x^n,y^n)) \nonumber \\
&\quad + \sqrt{\frac{2}{\ln(2)}}\sqrt{D(P(x^{n+1},y^{n+1})\| Q(x^{n+1},y^{n+1}))/n}.\label{bound.l1repeat}
\end{align}

\subsubsection{$L_1$ convergence rates}

We apply Lemma~\ref{lemma.ctwredundancy} in Appendix A. Plugging~(\ref{eqn.ctwredundancybounds2}) of Lemma~\ref{lemma.ctwredundancy} in~(\ref{bound.l1repeat}), we have
\begin{equation}
\mathbb{E} |C_n| = O((\log n)^{1/2} n^{-1/2}). \label{eqn.cnboundnew}
\end{equation}

\setcounter{equation}{120}

Combining~(\ref{eqn.cnboundnew}) with the $L_1$ convergence rates of $D_n$ shown in Lemma~\ref{lemma.causalaep} in Appendix~A, we have
\begin{align}
\mathbb{E}|\hat{H}_1 - \rH(\bY \| \bX)| &\leq \mathbb{E}|C_n| + \mathbb{E}|D_n| \\
&= O(n^{-1/2}\log n),
\end{align}
then we know the convergence rates in Proposition~\ref{cor.h2} hold as follows
\begin{equation}
\mathbb{E}\left |\hat{I}_1(X^n \to Y^n) - \rI(\bX\to\bY)\right | =  O(n^{-1/2}\log n).
\end{equation}

\subsubsection{Almost sure convergence rates}
We look at the almost sure convergence rates of $C_n$ (\ref{eqn.cnrepeat}) at first. We know the probability of event $\mathcal{A}_{n,\epsilon}$ defined in (\ref{eqn.typicallower}) is bounded as
\begin{equation}
\mathbb{P}(\mathcal{A}_{n,\epsilon}) \leq 2^{-n\epsilon}. \label{probability}
\end{equation}
For any fixed $\delta'>\delta>0$, taking $\epsilon = n^{-1+\delta}$ in (\ref{eqn.typicallower}), we see $\mathcal{A}_{n,\epsilon}$ is equal to the set
\begin{equation}
\{(x^n,y^n): n^{1-\delta'}\left(\hat{H}_1 + \frac{1}{n}\log P(y^n \| x^n)\right)\leq -n^{\delta - \delta'}\}.
\end{equation}
Note that
\begin{equation}
\sum_{n = 1}^\infty \mathbb{P}(\mathcal{A}_{n,\epsilon}) \leq \sum_{n = 1}^\infty 2^{-n^{\delta}} <\infty.
\end{equation}
By the Borel-Cantelli lemma, since $n^{\delta - \delta'}$ goes to zero as $n\to \infty$, we proved that
\begin{equation}
\liminf\limits_{n\to \infty} n^{1-\delta'}\left(\hat{H}_1 + \frac{1}{n}\log P(y^n\|x^n)\right) \geq 0 \quad \textrm{$P$-a.s.} \label{eqn.asfirstpart}
\end{equation}

In order to get an inequality of the reverse direction, dividing (\ref{eqn.expansion}) by $n^{-1+\delta'}$, we have
\begin{align}
& \quad n^{1-\delta'}\left(\hat{H}_1 + \frac{1}{n}\log P(Y^n\| X^n)\right) \\
& \qquad = n^{1-\delta'}\left( \frac{1}{n}\log \frac{P(Y^n,X^n)}{Q(Y^n,X^n)} \right) \nonumber \\
& \qquad  \quad -n^{1-\delta'}\left(\frac{1}{n}\log \frac{P(X^n\| Y^{n-1})}{Q(X^n\| Y^{n-1})}\right). \label{eqn.reverseexpansion}
\end{align}

By the pointwise redundancy of the CTW algorithm restated in Lemma~\ref{lemma.ctwredundancy} in Appendix A, we know
\begin{equation}
\limsup\limits_{n\to \infty}\frac{1}{\log n} \log \frac{P(Y^n,X^n)}{Q(Y^n,X^n)}  \leq 1\quad \textrm{$P$-a.s.} \label{eqn.rate1}
\end{equation}
then we have
\begin{equation}\label{eqn.secondpartfirst}
\limsup\limits_{n\to \infty}n^{1-\delta'}\left( \frac{1}{n}\log \frac{P(Y^n,X^n)}{Q(Y^n,X^n)} \right) \leq 0\quad \textrm{$P$-a.s.}
\end{equation}
For the second term on the right hand side of (\ref{eqn.reverseexpansion}), following similar argument applied to show (\ref{eqn.asfirstpart}), we know
\begin{equation}\label{eqn.secondpartsecond}
\limsup\limits_{n\to \infty}-n^{1-\delta'}\left(\frac{1}{n}\log \frac{P(X^n\| Y^{n-1})}{Q(X^n\| Y^{n-1})}\right) \leq 0\quad \textrm{$P$-a.s.}
\end{equation}
From (\ref{eqn.secondpartfirst}) and (\ref{eqn.secondpartsecond}), we obtain
\begin{equation}
\limsup\limits_{n\to \infty} n^{1-\delta'}\left(\hat{H}_1 + \frac{1}{n}\log P(Y^n\| X^n)\right) \leq 0\quad \textrm{$P$-a.s.} \label{eqn.assecondpart}
\end{equation}
Combining (\ref{eqn.asfirstpart}) and (\ref{eqn.assecondpart}) together, we know $\forall \delta'>0$,
\begin{equation}
\lim\limits_{n\to \infty} \hat{H}_1 + \frac{1}{n}\log P(Y^n\| X^n) = o(n^{-1+\delta'})\quad \textrm{$P$-a.s.}\label{ctw.asrate}
\end{equation}

Putting (\ref{ctw.asrate}) and the almost sure convergence rates of $D_n$ shown in Lemma~\ref{lemma.causalaep} in Appendix A together, we know $\forall \epsilon>0$,
\begin{equation*}
\hat{I}_1(X^n \to Y^n) - \rI(\bX\to\bY) = o(n^{-1/2}(\log n)^{5/2 + \epsilon}).\quad \textrm{$P$-a.s.}
\end{equation*}

\subsection{Proof of Theorem~\ref{thm.main}}\label{app.thm.main}
It suffices to show the convergence properties of $\hat{H}_2$. We decompose
\begin{equation}
\hat{H}_2(Y^n \| X^n) - \rH(\bY \| \bX) = A_n + B_n,
\end{equation}
where
\begin{align}
A_n & = \frac{1}{n}\sum_{k = 1}^n f(P(x_{k+1},y_{k+1}|X^{k},Y^{k}))- \rH(\bY \| \bX) \label{eqn.anfirstdefine}  \\
B_n & = \hat{H}_2(Y^n \| X^n) - \frac{1}{n}\sum_{k = 1}^n f(P(x_{k+1},y_{k+1}|X^{k},Y^{k})). \label{eqn.bnfirstdefine}
\end{align}
Define $g_k(\bX,\bY) \triangleq f( P(x_1,y_1|X_{-k}^{0},Y_{-k}^{0}))$ for a jointly stationary and ergodic process $(\bX,\bY)$. Note that, by martingale convergence \cite{Breiman1992}, $g_k(\bX,\bY)\to g(\bX,\bY),\textrm{$P$-a.s.}$ where $g(\bX,\bY) = f( P(x_1,y_1|X_{-\infty}^{0},Y_{-\infty}^{0}) )$. Noting further that $\E g(\bX, \bY) = \rH(\bY|| \bX)$ and $\forall k, g_k$ are bounded, we can apply Lemma~\ref{lemma.seconv} in Appendix A and get the following result:
\begin{equation}
\lim_{n\to\infty} A_n = 0\quad\textrm{$P$-a.s. and in $L_1$}. \label{cor.convergence}
\end{equation}
Then we deal with $B_n$ defined in (\ref{eqn.bnfirstdefine}) from (\ref{eqn.hpqfirst}) to (\ref{eq.HQ}), where fixing an arbitrary $\epsilon>0$,
\begin{itemize}
\item (\ref{eqn.hpqa}) follows by Lemma~\ref{lemma.unifconv} in Appendix A,
\item (\ref{eqn.hpqb}) follow by Pinsker's inequality,
\item (\ref{eqn.hpqc}) and (\ref{eqn.hpqd}) follow by the concavity of $\sqrt{\cdot}$,
\item (\ref{eq.HQ}) follows by the chain rule for relative entropy.
\end{itemize}

\begin{figure*}[b!]
\setcounter{equation}{154}
\hrulefill
\begin{align}
\E |B_n| &=\E\left|\frac{1}{n}\sum_{k=1}^n\left( \funcf{Q(x_{k+1},y_{k+1}|X^{k},Y^{k})}-\funcf{P(x_{k+1},y_{k+1}|X^{k},Y^{k})}\right)\right|\label{eqn.hpqfirst}\\
&\hspace{-0.21cm}\leq\frac{1}{n} \E\sum_{k=1}^n\left| \funcf{Q(x_{k+1},y_{k+1}|X^{k},Y^{k})}-\funcf{P(x_{k+1},y_{k+1}|X^{k},Y^{k})} \right|\\
&\hspace{-0.21cm}\leq
\frac{1}{n} \sum_{k=1}^n\E \left(\epsilon+K_\epsilon\norm{Q(x_{k+1},y_{k+1}|X^{k},Y^{k})-P(x_{k+1},y_{k+1}|X^{k},Y^{k})}_1\right)\label{eqn.hpqa}\\
&\hspace{-0.21cm}\leq
\frac{K_\epsilon}{n}\sum_{k=1}^n\E\left[\sqrt{2\ln(2)D\left(P(x_{k+1},y_{k+1}|X^{k},Y^{k}) || Q(x_{k+1},y_{k+1}|X^{k},Y^{k})\right)}\right]+ \epsilon\label{eqn.hpqb}\\
&\hspace{-0.21cm}\leq
\frac{K_\epsilon}{n}\sum_{k=1}^n\sqrt{2\ln(2)\E\left[ D\left(P(x_{k+1},y_{k+1}|X^{k},Y^{k}) || Q(x_{k+1},y_{k+1}|X^{k},Y^{k})\right)\right]}+ \epsilon \label{eqn.hpqc}\\
&\hspace{-0.21cm}= \epsilon + \frac{K_\epsilon}{n}\sum_{k=1}^n\sqrt{2\ln(2)\mathbb{E}D\left(P(x_{k+1},y_{k+1}|X^{k},Y^{k}) || Q(x_{k+1},y_{k+1}|X^{k},Y^{k}) \right)}\\
&\hspace{-0.21cm}\leq
\epsilon +K_\epsilon \sqrt{\frac{2\ln(2)}{n}}\times\sqrt{\sum_{k=1}^n                             \mathbb{E}D\left(P(x_{k+1},y_{k+1}|X^{k},Y^{k})||Q(x_{k+1},y_{k+1}|X^{k},Y^{k})\right)}\label{eqn.hpqd}\\  \displaybreak
&\hspace{-0.21cm}= \epsilon + K_\epsilon\sqrt{\frac{2\ln(2)}{n}D\left(P(x^{n+1},y^{n+1})||Q(x^{n+1},y^{n+1})\right)} \label{eq.HQ}
\end{align}
\end{figure*}

\setcounter{equation}{138}

We continue to bound
\begin{align}
&\lim\limits_{n\to\infty}\E\left|\hat{H}_2(Y^n \| X^n)-\rH(\bY \| \bX)\right|\\
& \qquad \leq \lim\limits_{n\to\infty}\E\left|A_n\right| +\lim\limits_{n\to\infty}\E\left|B_n\right|\\
&\qquad = \lim\limits_{n\to\infty}\E\left|B_n\right|\label{eqn.hbarf}\\
& \qquad \leq \epsilon + \lim\limits_{n\to\infty}K_\epsilon\sqrt{\frac{2\ln(2)}{n}D\left(P(x^{n+1},y^{n+1})||Q(x^{n+1},y^{n+1})\right)}\label{eqn.hbarg}\\
& \qquad = \epsilon \label{eqn.hbarh}
\end{align}
where (\ref{eqn.hbarf}) follows by (\ref{cor.convergence}), (\ref{eqn.hbarg}) follows by (\ref{eq.HQ}), (\ref{eqn.hbarh}) follows by Definition~\ref{def.Q}.  Now we can use the arbitrariness of $\epsilon$ to complete the proof.

\subsection{Proof of Proposition~\ref{cor.hs}}\label{app.cor.hs}

It suffices to show the convergence properties of $\hat{H}_2$.
\subsubsection{Almost sure convergence}
For stationary ergodic process $(\bX, \bY)$, let
\begin{align}
g_k(\bX, \bY) & = f(Q(x_0,y_0|X^{-1}_{-k})) \\
g(\bX, \bY) & = f(P(x_0,y_0|X_{-\infty}^{-1},Y_{-\infty}^{-1})),
\end{align}
by Lemma~\ref{lemma.as} in Appendix A,
\begin{equation}
 \lim\limits_{k\to \infty}g_k(\bX, \bY) - g(\bX, \bY) = 0 \quad \textrm{$P$-a.s.}
\end{equation}
Since $\mathbb{E}[\sup\limits_{k}|g_k|]\leq \log |\cY|$, by Lemma~\ref{lemma.seconv} in Appendix A,
\begin{equation}
 \lim\limits_{n\to \infty} \frac{1}{n} \sum_{k = 1}^n g_k(T^k(\bX,\bY)) =  \lim\limits_{n\to \infty}\hat{H}_2 = \rH(\bY \| \bX),
\end{equation}
which justifies the almost sure convergence of $\hat{H}_2$.

\subsubsection{$L_1$ convergence rates}
For convenience, we restate the definitions of $A_n$ and $B_n$ as follows
\begin{align}
A_n & = \frac{1}{n}\sum_{k = 1}^n f(P(x_{k+1},y_{k+1}|X^{k},Y^{k}))- \rH(\bY \| \bX) \\
B_n & =  \hat{H}_2(Y^n \| X^n)- \frac{1}{n}\sum_{k = 1}^n f(P(x_{k+1},y_{k+1}|X^{k},Y^{k})).
\end{align}
Letting $V_k$ be $f(P(x_{k+1},y_{k+1}|X^{k},Y^{k}))- \rH(\bY \| \bX)$, $V$ be $\log |\mathcal{Y}|$, and applying Lemma~\ref{lemma.mixing} in Appendix A, we know
\begin{equation}
\mathbb{E}|A_n| \leq \sqrt{\mathbb{E}A^2_n} = O(n^{-1/2}). \label{eqn.boundan}
\end{equation}
Then we bound $\mathbb{E}|B_n|$ from (\ref{eqn.bnfirst}) to (\ref{eqn.zhaoproof}). 
\begin{itemize}
\item (\ref{eqn.bna}) is an application of Lemma~\ref{lemma.as} and Lemma~\ref{lemma.diff} in Appendix A. Indeed, Lemma~\ref{lemma.as} guarantees that when $n\to\infty$, the $\ell_1$ norm of the difference of $P(x_{k+1},y_{k+1}|X^k,Y^k)$ and $Q(x_{k+1},y_{k+1}|X^k,Y^k)$ will be small enough so that Lemma~\ref{lemma.diff} can be applied.
\item (\ref{eqn.bnb}) follows by Pinsker's inequality and the fact that function $t\log(1/t)$ is increasing for small $t$.
\item (\ref{eqn.bnc}) and (\ref{eqn.zhaoproof}) are by the concavity of $\sqrt{t}\log(1/\sqrt{t})$ and the chain rule for relative entropy.
\end{itemize}

\begin{figure*}[b!]
\setcounter{equation}{177}
\hrulefill
\begin{align}
\mathbb{E}\left |B_n\right | & = \mathbb{E}\left |\frac{1}{n}\sum_{k = 1}^n \left (f(Q(x_{k+1},y_{k+1}|X^{k},Y^{k}))-f(P(x_{k+1},y_{k+1}|X^{k},Y^{k}))\right)\right | \label{eqn.bnfirst} \\
& \leq \frac{1}{n}\mathbb{E} \sum_{k = 1}^n \left |f(Q(x_{k+1},y_{k+1}|X^{k},Y^{k}))-f(P(x_{k+1},y_{k+1}|X^{k},Y^{k}))\right | \\
& \leq  \frac{1}{n}\mathbb{E}\sum_{k = 1}^n 2\| P(x_{k+1},y_{k+1}|X^{k},Y^{k}) - Q(x_{k+1},y_{k+1}|X^{k},Y^{k})\|_1 \nonumber \\
& \qquad \times \log\frac{|\mathcal{X}||\mathcal{Y}|}{\| P(x_{k+1},y_{k+1}|X^{k},Y^{k}) - Q(x_{k+1},y_{k+1}|X^{k},Y^{k})\|_1} \label{eqn.bna}\\
& \leq  \frac{1}{n}\mathbb{E}\sum_{k = 1}^n 2\sqrt{2\ln(2)D(P(x_{k+1},y_{k+1}|X^{k},Y^{k})\| Q(x_{k+1},y_{k+1}|X^{k},Y^{k}))}\nonumber\\
& \qquad \times \log\frac{|\mathcal{X}||\mathcal{Y}|}{\sqrt{2\ln(2)D(P(x_{k+1},y_{k+1}|X^{k},Y^{k}) \| Q(x_{k+1},y_{k+1}|X^{k},Y^{k}))}} \label{eqn.bnb}\\
& \leq  \frac{1}{n}\sum_{k = 1}^n 2\sqrt{2\ln(2)\mathbb{E}D(P(x_{k+1},y_{k+1}|X^{k},Y^{k})\| Q(x_{k+1},y_{k+1}|X^{k},Y^{k}))}\nonumber \\
& \qquad \times \log\frac{|\mathcal{X}||\mathcal{Y}|}{\sqrt{2\ln(2)\mathbb{E}D(P(x_{k+1},y_{k+1}|X^{k},Y^{k}) \| Q(x_{k+1},y_{k+1}|X^{k},Y^{k}))}} \label{eqn.bnc}\\
& \leq  2\sqrt{2\ln(2)D(P(x^{n+1},y^{n+1})\| Q(x^{n+1},y^{n+1}))/n}\log \frac{|\mathcal{X}||\mathcal{Y}|}{\sqrt{2\ln(2)D(P(x^{n+1},y^{n+1})\| Q(x^{n+1},y^{n+1}))/n}}\label{eqn.zhaoproof}
\end{align}
\end{figure*}

\setcounter{equation}{150}

Because of the monotonicity of $\sqrt{t}\log(1/\sqrt{t})$ when $t\approx 0$, we can plug in the redundancy bounds of the CTW algorithm in Lemma~\ref{lemma.ctwredundancy} in Appendix A, i.e., (\ref{eqn.ctwredundancybounds2}) into (\ref{eqn.zhaoproof}), then have
\begin{equation}
\mathbb{E} |B_n| = O  (n^{-1/2}(\log n)^{3/2}). \label{eqn.boundonbn}
\end{equation}
Combining (\ref{eqn.boundonbn}) with (\ref{eqn.boundan}), we proved Proposition~\ref{cor.hs}.

\subsection{Proof of Proposition~\ref{thm.minimax}}\label{app.thm.minimax}

We rephrase a general lemma showing minimax lower bounds:
\begin{lemma}[\!\!{{\cite[Theorem 2.2, Page 90]{Tsybakov2008}}}] \label{lemma.tsybakov}
Let $\mathcal{F}$ be a class of models, and suppose we have observations $Z$ distributed according to $P_f$,$f\in \mathcal{F}$. Let $d(\hat{f},f)$ be the performance measure of the estimator $\hat{f}(Z)$ relative to the true model $f$. Assume
also $d(\cdot,\cdot)$ is a semi-distance, i.e., it satisfies
\begin{enumerate}
\item $d(f,g) = d(g,f)\geq 0,$
\item $d(f,f) = 0$,
\item $d(f,g)\leq d(h,f) + d(h,g)$.
\end{enumerate}
Let $f_0,f_1\in \mathcal{F}$ satisfy $d(f_0,f_1)\geq 2s>0$, where $s$ is fixed. Then
\begin{align}
\inf\limits_{\hat{f}} \sup\limits_{f\in \mathcal{F}}P_f(d(\hat{f},f)\geq s) & \geq  \inf\limits_{\hat{f}} \max\limits_{j\in \{0,1\}}P_{f_j}(d(\hat{f},f_j)\geq s) \\
& \geq  \frac{1}{4}\exp(-D(P_{f_1}\| P_{f_0})).
\end{align}
\end{lemma}

In this proof, $\mathcal{F}$ in Lemma~\ref{lemma.tsybakov} is taken to be $\mathscr{P}(\bX,\bY)$. Denote the binary entropy as $H_b(p) = -p\log p  - (1-p)\log (1-p)$ and the class of i.i.d. processes as $\mathcal{M}_0$. Since
\begin{equation}
H_b'(p) = \log \frac{1-p}{p},
\end{equation}
and $H_b'(p)$ is decreasing in interval $[2/8,3/8]$, we know

\setcounter{equation}{162}

\begin{lemma}\label{lemma.bern_pq}
$\forall p,q\in [2/8,3/8]$, we have
\begin{equation}
|H_b(p)-H_b(q)|\geq \log (5/3)|p-q|.
\end{equation}
\end{lemma}

We also show a lemma bounding the divergence between two Bernoulli pmfs.
\begin{lemma} \label{lemma.bernoulli}
Let $P$ and $Q$ be Bernoulli pmfs with parameters, respectively, 1/2-$p$ and 1/2-$q$. If $|p|,|q|\leq 1/4$, then $D(P\| Q)\leq 8(p-q)^2$.
\end{lemma}

Lemma~\ref{lemma.bernoulli} can be verified as follows:
\begin{align}
D(P \| Q) & = (1/2-p)\log \frac{1/2-p}{1/2-q} + (1/2+p)\log \frac{1/2+p}{1/2+q} \\
& = (1/2-p)\log \left( 1+ \frac{q-p}{1/2-q}\right) \nonumber \\
& \quad + (1/2+p) \log \left( 1+ \frac{p-q}{1/2+q}\right) \\
& \leq \frac{1}{\ln(2)}\left( (1/2-p)\frac{q-p}{1/2-q} + (1/2+p)\frac{p-q}{1/2+q} \right)\\
& = \frac{1}{\ln(2)}\frac{(p-q)^2}{1/4-q^2} \\
& \leq 8(p-q)^2,
\end{align}
where the first inequality holds because $\log(1+x) \leq x/\ln(2), \forall x>-1$, and the second inequality holds because $|q|\leq 1/4$.

Taking the observations model as $X_i\stackrel{\mathrm{i.i.d.}}{\sim}\mathrm{Bern}(q)$, $Y_i = X_i$, then we have $\DIR{\bX}{\bY} = H(X)$. Assuming under model $f_0, q = q_0 = 1/4$, under model $f_1$, $q = q_1 = 1/4 + 1/\sqrt{n}$, and $n\geq 64$. Let $\hat{I}_n$ be an arbitrary estimator of $\DIR{\bX}{\bY}$ based on $(X_1^n, Y_1^n)$, $d(x,y) = |x-y|$, we have
\begin{equation}
d(H_b(q_0),H_b(q_1))\geq \log(5/3)|q_0-q_1| =  \log(5/3)/\sqrt{n}.
\end{equation}
Then we take $s = \log(5/3)/(2\sqrt{n})$ to satisfy the assumption of Lemma~\ref{lemma.tsybakov}. For brevity, here we denote $\DIR{\bX}{\bY}$ as $I$. By Lemma~\ref{lemma.tsybakov},
\begin{align}
\inf\limits_{\hat{I}_n} \sup\limits_{\mathcal{M}_0}P_f(d(\hat{I}_n,I)\geq s)
 & \geq \inf\limits_{\hat{I}_n} \max\limits_{j\in \{0,1\}}P_{f_j}(d(\hat{I}_n,H_b(q_j))\geq s)\\
& \geq  \frac{1}{4}\exp(-D(P_{f_1}\| P_{f_0})).
\end{align}
Then we bound $D(P_{f_1}\| P_{f_0})$:
\begin{align}
D(P_{f_1}\| P_{f_0}) & = n\mathbb{E}_{f_1}\left[\log \frac{P_{f_1}(X_1,Y_1)}{P_{f_0}(X_1,Y_1)}\right] \\
& \leq 8n(q_0-q_1)^2 \\
& = 8.
\end{align}
Thus we have
\begin{equation}
\inf\limits_{\hat{I}_n} \sup\limits_{\mathcal{M}_0}P_f(d(\hat{I}_n,I)\geq s) \geq \frac{1}{4}e^{-8}.
\end{equation}
Using Markov's inequality,
\begin{align}
\inf\limits_{\hat{I}_n} \sup\limits_{\mathscr{P}(\bX,\bY)}\mathbb{E}|\hat{I}_n-I| & \geq \inf\limits_{\hat{I}_n} \sup\limits_{\mathcal{M}_0}\mathbb{E}|\hat{I}_n-I| \\
& \geq \frac{1}{4}e^{-8}s = \frac{1}{8}e^{-8}\log(5/3)\frac{1}{\sqrt{n}}.
\end{align}

\setcounter{equation}{183}

\subsection{Proof of Theorem~\ref{thm.hybrid}}\label{app.thm.hybrid}
We decompose
\begin{align}
\hat{I}_3 & =  \frac{1}{n}\sum_{i= 1}^n\sum_{y_i} Q(y_i|X^i,Y^{i-1})\log \frac{1}{Q(y_i|Y^{i-1})} \nonumber\\
& \quad -  \frac{1}{n}\sum_{i= 1}^n\sum_{y_i} Q(y_i|X^i,Y^{i-1})\log \frac{1}{Q(y_i|X^i,Y^{i-1})}. \label{eqn.decom1}
\end{align}
Following the proof of almost sure and $L_1$ convergence of $\hat{H}_2$ in that of Proposition~\ref{cor.hs}, we can show that the second term on the right hand side of (\ref{eqn.decom1}) converges to $\rH(\bY \| \bX)$ almost surely and in $L_1$ under the conditions of Theorem~\ref{thm.hybrid}. Denote the first term on the right hand size of (\ref{eqn.decom1}) as
\begin{equation}
F_n = \frac{1}{n}\sum_{i= 1}^n\sum_{y_i} Q(y_i|X^i,Y^{i-1})\log \frac{1}{Q(y_i|Y^{i-1})}.
\end{equation}
Then it suffices to show the almost sure and $L_1$ convergence of $F_n$ to $\rH(\bY)$. Decompose $F_n-\rH(\bY)$ as
\begin{equation*}
F_n-\rH(\bY) = R_n + S_n,
\end{equation*}
where
\begin{align} \allowdisplaybreaks
R_n & = \frac{1}{n}\sum_{i = 1}^n \sum_{y_i} P(y_i|X^i,Y^{i-1})\log P(y_i|Y^{i-1})\nonumber \\
&\quad - \frac{1}{n}\sum_{i = 1}^n\sum_{y_i}Q(y_i|X^i,Y^{i-1})\log Q(y_i|Y^{i-1}) \label{eqn.rnthm3}\\
S_n & = -\frac{1}{n}\sum_{i = 1}^n \sum_{y_i} P(y_i|X^i,Y^{i-1})\log P(y_i|Y^{i-1}) -\rH(\bY). \label{eqn.snthm3}
\end{align}

\subsubsection{Almost sure convergence}
Express $R_n$ as $\frac{1}{n}\sum_{i = 1}^n Z_i$, where
\begin{align}
Z_i & = -\sum_{y_i}Q(y_i|X^i,Y^{i-1})\log Q(y_i|Y^{i-1})\nonumber \\
& \quad + \sum_{y_i}P(y_i|X^i,Y^{i-1})\log P(y_i|Y^{i-1}).
\end{align}

According to Lemma~\ref{lemma.as} in Appendix A, the CTW probability assignments, $Q(y_i|X^i,Y^{i-1})$ and $Q(y_i|Y^{i-1})$ both converge almost surely to the true probability $P(y_i|X^i,Y^{i-1})$ and $P(y_i|Y^{i-1})$. Therefore, 
\begin{equation}
\lim\limits_{i\to \infty}Z_i = 0\quad \textrm{$P$-a.s.}
\end{equation}
Then we know the Ces\'{a}ro mean of $\{Z_i\}_{i = 1}^n$ also converges to zero almost surely, i.e.,
\begin{equation}
\lim\limits_{n\to \infty}R_n = \lim\limits_{n\to \infty}\frac{1}{n}\sum_{i = 1}^n Z_i =0\quad \textrm{$P$-a.s.}
\end{equation}
Now we show $S_n$ converges to zero almost surely, which is implied by Birkhoff's ergodic theorem.

\subsubsection{$L_1$ convergence}
We express $R_n$ in another form in (\ref{eqn.rnanother}), and bound $\mathbb{E}|R_n|$ from (\ref{eqn.rnfirst}) to (\ref{eqn.float2}), where
\begin{itemize}
\item The first part of (\ref{eqn.rna}) is derived by (\ref{eqn.pinsker}), and the second part of (\ref{eqn.rna}) is implied by the fact that the CTW probability assignment is lower bounded (\ref{eqn.bounded});
\item (\ref{eqn.rnb}) follows by Pinsker's inequality,
\item (\ref{eqn.rnc}) follows by data-processing inequality,
\item (\ref{eqn.float2}) follows by the chain rule of relative entropy and concavity of $\sqrt{\cdot}$.
\end{itemize}

\begin{figure*}[b!]
\setcounter{equation}{204}

\hrulefill
\begin{align}
R_n & = \frac{1}{n}\sum_{i = 1}^n \sum_{y_i} P(y_i|X^i,Y^{i-1})\log \frac{P(y_i|Y^{i-1})}{Q(y_i|Y^{i-1})} \nonumber\\
& \quad  + \frac{1}{n}\sum_{i = 1}^n\sum_{y_i}\left (P(y_i|X^i,Y^{i-1})-Q(y_i|X^i,Y^{i-1})\right )\log Q(y_i|Y^{i-1}), \label{eqn.rnanother} \\
\mathbb{E}|R_n| & \leq \frac{1}{n}\sum_{i = 1}^n \mathbb{E}\left |\sum_{y_i} P(y_i|X^i,Y^{i-1})\log \frac{P(y_i|Y^{i-1})}{Q(y_i|Y^{i-1})}\right|  \nonumber \\
& \quad +  \frac{1}{n}\sum_{i = 1}^n\mathbb{E}\left |\sum_{y_i}\left (P(y_i|X^i,Y^{i-1})-Q(y_i|X^i,Y^{i-1})\right )\log Q(y_i|Y^{i-1})\right |\label{eqn.rnfirst}\\
& \leq \frac{1}{n}\sum_{i = 1}^n \mathbb{E}\left [\sum_{y_i} P(y_i|X^i,Y^{i-1})\left|\log \frac{P(y_i|Y^{i-1})}{Q(y_i|Y^{i-1})}\right|\right]  \nonumber \\
& \quad +  \frac{1}{n}\sum_{i = 1}^n\mathbb{E}\left |\sum_{y_i}\left (P(y_i|X^i,Y^{i-1})-Q(y_i|X^i,Y^{i-1})\right )\log Q(y_i|Y^{i-1})\right |\label{eqn.rnfirst_new}\\
& \leq \frac{1}{n}\sum_{i = 1}^n \mathbb{E}\left [\sum_{y_i} P(y_i|Y^{i-1})\left|\log \frac{P(y_i|Y^{i-1})}{Q(y_i|Y^{i-1})}\right|\right]  \nonumber \\
& \quad+  \frac{1}{n}\sum_{i = 1}^n \mathbb{E}\left [\sum_{y_i}\log\frac{1}{Q(y_i|Y^{i-1})}\left |P(y_i|X^i,Y^{i-1})-Q(y_i|X^i,Y^{i-1})\right |\right ]\\
& \leq  \frac{1}{n}\sum_{i = 1}^n \left (\mathbb{E}D(P(y_i|Y^{i-1})\| Q(y_i|Y^{i-1}))+ \sqrt{\frac{2}{\ln(2)}}\sqrt{\mathbb{E}D(P(y_i|Y^{i-1})\| Q(y_i|Y^{i-1}))}\right) \nonumber \\
& \quad + \frac{1}{n}\sum_{i = 1}^n \mathbb{E}\left [\log (2i+|\mathcal{Y}|)\sum_{y_i}|P(y_i|X^i,Y^{i-1})-Q(y_i|X^i,Y^{i-1})|\right] \label{eqn.rna}\\
& \leq  \frac{1}{n}\sum_{i = 1}^n \left (\mathbb{E}D(P(y_i|Y^{i-1})\| Q(y_i|Y^{i-1}))+ \sqrt{\frac{2}{\ln(2)}}\sqrt{\mathbb{E}D(P(y_i|Y^{i-1})\| Q(y_i|Y^{i-1}))}\right) \nonumber \\
& \quad + \frac{1}{n}\sum_{i = 1}^n \log (2i+|\mathcal{Y}|)\sqrt{2\ln(2) \mathbb{E}D(P(y_i|X^i,Y^{i-1})\| Q(y_i|X^i,Y^{i-1}))}\label{eqn.rnb}\\
& \leq  \frac{1}{n}\sum_{i = 1}^n \left (\mathbb{E}D(P(y_i|Y^{i-1})\| Q(y_i|Y^{i-1}))+ \sqrt{\frac{2}{\ln(2)}}\sqrt{\mathbb{E}D(P(y_i|Y^{i-1})\| Q(y_i|Y^{i-1}))}\right) \nonumber \\
& \quad + \frac{1}{n}\sum_{i = 1}^n \log (2i+|\mathcal{Y}|)\sqrt{2\ln(2) \mathbb{E}D(P(x_i,y_i|X^i,Y^{i-1})\| Q(x_i,y_i|X^i,Y^{i-1}))}\label{eqn.rnc}\\
& \leq  \frac{1}{n}D(P(y^n)\| Q(y^n))+ \sqrt{\frac{2}{\ln(2)}\frac{D(P(y^n)\| Q(y^n))}{n}} \nonumber  \\
& \quad + \log (2n+|\mathcal{Y}|)\sqrt{\frac{2\ln(2)D(P(x^n,y^n)\| Q(x^n,y^n))}{n}},\label{eqn.float2}
\end{align}
\end{figure*}

\setcounter{equation}{190}

After applying Lemma~\ref{lemma.ctwredundancy} in Appendix A, we know $R_n$ converges to zero in $L_1$. By Birkhoff's ergodic theorem, we know the convergence of $S_n$ is also in $L_1$, which completes the proof of $L_1$ convergence.

\subsection{Proof of Theorem~\ref{thm.diver}}\label{app.thm.diver}

We decompose $\hat{I}_4$
\begin{equation}
\hat{I}_4 = G_n - \hat{H}_2,
\end{equation}
where $\hat{H}_2$ is the estimator for $\rH(\bY \| \bX)$ in $\hat{I}_2$, $G_n$ is defined as
\begin{equation}
G_n = \frac{1}{n}\sum_{i= 1}^n\sum_{(x_{i+1},y_{i+1})} Q(x_{i+1},y_{i+1}|X^i,Y^i)\log \frac{1}{Q(y_{i+1}|Y^i)}.
\end{equation}

Since $G_n$ is in similar form as $F_n$, we can follow corresponding steps in the proof of Theorem~\ref{thm.hybrid} to establish Theorem~\ref{thm.diver} analogously.

\section{Proofs of Technical Lemmas}

\subsection{Proof of Lemma~\ref{lemma.causalaep}}

\subsubsection{General stationary ergodic processes}
The convergence holds almost surely
by the Shannon--McMillan--Breiman theorem for causally conditional entropy rate (see, for example, \cite{Venkataramanan--Pradhan2007}). We now prove the AEP also holds in $L_1$.

Denote
\begin{align}
A_n & = -\frac{1}{n}\log P(Y^n \| X^n) \\
B_n & = -\frac{1}{n}\log P(Y^n \| X^n, X^0_{-\infty}, Y^0_{-\infty}) \\
C_n & = B_n - A_n,
\end{align}
where $P(Y^n \| X^n, X^0_{-\infty},Y^0_{-\infty}) = \prod_{i = 1}^n P(Y_i|X^i_{-\infty},Y^{i-1}_{-\infty})$. Our goal is to show that $\mathbb{E}|A_n - \rH (\bY \| \bX)|$ converges to zero when $n \to \infty$. 

Note that \begin{align}
\mathbb{E}A_n & = \frac{1}{n} \sum_{i = 1}^n H(Y_i|Y^{i-1},X^i),\\
\mathbb{E}B_n & = \rH (\bY \| \bX).
\end{align}

By stationarity of $(\mathbf{X, Y})$ and conditioning reduces entropy, we know $H(Y_i|Y^{i-1},X^i)$ is a nonnegative, nonincreasing sequence in $i$, and further, it converges to $\rH (\bY \| \bX)$. Since $\mathbb{E}A_n$ is the Ces\'aro mean of sequence $\{H(Y_i|Y^{i-1},X^i)\}_{i = 1}^n$, it follows that $\mathbb{E}A_n$ converges to $\rH (\bY \| \bX)$ as $n \to \infty$. Thus,
\begin{equation} \label{eqn.cnconverge}
\lim\limits_{n \to \infty}\mathbb{E}C_n = 0.
\end{equation}
We have
\begin{align}
\mathbb{E}|A_n - \rH (\bY \| \bX)| & = \mathbb{E}|A_n -\mathbb{E}B_n |  \\
& \leq \mathbb{E}|C_n| + \mathbb{E}|B_n - \mathbb{E}B_n|.\label{eqn.aepfirstbound}
\end{align}
By Birkhoff's ergodic theorem, $\mathbb{E}|B_n - \mathbb{E}B_n|$ converges to zero when $n \to \infty$. It now suffices to show that $\lim\limits_{n \to \infty}\mathbb{E}|C_n| = 0$. Denote the CDF of random variable $C_n$ as $F_n(x)$, then we have
\begin{align}
\mathbb{E}|C_n| & = -\mathbb{E}C_n + 2\int_0^{\infty} xdF_n(x),  \\
& = -\mathbb{E}C_n + 2\int_0^{\infty} P(C_n >x)dx, \label{eqn.aepkimbound}
\end{align}
where the second step follows by integration by parts and the fact that $1-F_n(x) = P(C_n >x)$. Let $B(X^0_{-\infty},Y^0_{-\infty}) \triangleq \{(x^n, y^n): P(x^n, y^n|X^0_{-\infty}, Y^0_{-\infty})>0\} $, we have (\ref{eqn.appendixcover}), then by Markov's inequality, we have
\begin{equation}
P\left (\frac{P(Y^n \| X^n)}{P(Y^n \| X^n, X^0_{-\infty},Y^0_{-\infty})} \geq t_n\right ) \leq \frac{1}{t_n},
\end{equation}
for arbitrary positive $t_n$.

\begin{figure*}[b!]

\setcounter{equation}{229}
\hrulefill
\begin{align}
\mathbb{E}\left [ \frac{P(Y^n \| X^n)}{P(Y^n \| X^n, X^0_{-\infty},Y^0_{-\infty})}\right ] & = \mathbb{E}\left [\mathbb{E}\left \{ \left.  \frac{P(Y^n \| X^n)}{P(Y^n \| X^n, X^0_{-\infty}, Y^0_{-\infty})} \right | X^0_{-\infty}, Y^0_{-\infty}\right \} \right ]  \\
& = \mathbb{E}\left[  \sum_{(x^n, y^n) \in B(X^0_{-\infty},Y^0_{-\infty}) } \frac{P(y^n \| x^n)}{P(y^n \| x^n, X^0_{-\infty}, Y^0_{-\infty})} P(x^n, y^n | X^0_{-\infty}, Y^0_{-\infty})         \right] \\
& = \mathbb{E}\left[ \sum_{(x^n, y^n) \in B(X^0_{-\infty},Y^0_{-\infty}) } P(y^n \|x^n)P(x^n \| y^{n-1}, X^0_{-\infty}, Y^0_{-\infty})    \right] \\
& \leq \sum_{(x^n,y^n)} P(y^n \|x^n)P(x^n \| y^{n-1}) \\
& = \sum_{(x^n,y^n)}  P(x^n, y^n) \\
& = 1. \label{eqn.appendixcover}
\end{align}
\end{figure*}

\setcounter{equation}{203}

Taking $t_n = 2^{nx}, x\geq 0$, we have
\begin{equation}
P\left (\frac{1}{n} \log \frac{P(Y^n \| X^n)}{P(Y^n \| X^n, X^0_{-\infty},Y^0_{-\infty})} \geq x\right ) \leq 2^{-nx}.
\end{equation}

\setcounter{equation}{212}

Equivalently,
\begin{equation} \label{eqn.aepcoverbound}
P(C_n >x) \leq 2^{-nx}.
\end{equation}
Plugging (\ref{eqn.aepcoverbound}) into (\ref{eqn.aepkimbound}), we have
\begin{align}
\mathbb{E}|C_n| & =  -\mathbb{E}C_n + 2\int_0^{\infty} P(C_n >x)dx \\
& \leq -\mathbb{E}C_n  + \frac{2}{n\ln (2)}.
\end{align}
By (\ref{eqn.cnconverge}), we know
\begin{equation}
\lim\limits_{n\to \infty} \mathbb{E}|C_n| = 0.
\end{equation}
By (\ref{eqn.aepfirstbound}), we know the AEP for causally conditional entropy holds in $L_1$.

\subsubsection{Irreducible aperiodic Markov processes}
We express $-\frac{1}{n}\log P(Y^n \| X^n) - \rH (\bY \| \bX)$ as
\begin{equation}
-\frac{1}{n}\log P(Y^n \| X^n) - \rH (\bY \| \bX) = \frac{1}{n}\sum_{i = 1}^n Z_i,
\end{equation}
where
\begin{equation}
Z_i = -\log P(Y_i|X^i_{i-m},Y^{i-1}_{i-m})-\rH(\bY \| \bX),
\end{equation}
and $m$ is the order of the Markov process $\mathbf{(X,Y)}$. Let
\begin{equation}
g_i = -\log P(Y_i|X^i_{i-m},Y^{i-1}_{i-m})
\end{equation}
and denote $\mathbb{E} g_i$ by $H$. Here $H$ does not depend on $i$ since the Markov process is stationary.

We decompose $Z_i$ as
\begin{equation}
Z_i = (g_i^L - H^L) + (g_i^{L'} - H^{L'}),
\end{equation}
where $g_i^L =g_i\mathbf{1}_{\{|g_i|\leq L\}}$, $g_i^{L'} = g_i - g_i^{L}$, $H^L=\mathbb{E}g_i^L$,
and $H^{L'} = \mathbb{E}g_i^{L'}= \rH(\bY \| \bX)-H^L$. We expand
\begin{align}
\mathbb{E}\left(\sum_{i = 1}^n Z_{i}\right)^2
& = \mathbb{E}\left(\sum_{i = 1}^n g_i^L - H^L\right)^2 + \mathbb{E}\left( \sum_{i = 1}^n g_i^{L'}-H^{L'}\right)^2  \nonumber\\
& \qquad+ 2\mathbb{E}\left(\sum_{i = 1}^n g_i^L - H^L\right)\left( \sum_{i = 1}^n g_i^{L'}-H^{L'}\right)
\label{eqn.fourterms}
\end{align}
and bound the three terms on the right hand side of~(\ref{eqn.fourterms}) separately.

For the first term, by Lemma~\ref{lemma.mixing} in Appendix A with $\mathbf{X} \leftarrow(\mathbf{X,Y})$, $V_i \leftarrow g_i^L-H^L$,
and $V \leftarrow L$, we have
\begin{equation}
\mathbb{E}\left(\sum_{i = 1}^n g_i^L - H^L\right)^2  = O(nL^2).
\end{equation}
For the second term, consider
\begin{equation}
\mathbb{E}\left( \sum_{i = 1}^n g_i^{L'}-H^{L'}\right)^2 \leq n^2\max\limits_{i}\mathbb{E}(g_i^{L'} - H^{L'})^2.
\end{equation}
Define
\begin{equation}
E_{i,K} = \{(x^i_{i-m},y^i_{i-m}):K \leq -\log P(y_i|x^i_{i-m},y^{i-1}_{i-m}) \leq K+1\},
\end{equation}
we have
\begin{align}
\mathbb{E}(g_i^{L'} - H^{L'})^2 & \leq \mathbb{E}(g_i^{L'})^2  \\
& \leq \sum_{K = L}^{\infty} \int_{E_{i,K}}(\log P(Y_i|X^i_{i-m},Y^{i-1}_{i-m}))^2d\mu \\
& \leq \sum_{K = L}^{\infty}|\mathcal{Y}|(K+1)^2 2^{-K}  \\
&  =  O(L^2 2^{-L}),
\end{align}
where the last inequality is an inequality developed by McMillan\cite{McMillan1953}, and the last step could be intuitively understood since the terms decay rapidly, the sum is dominated by the largest term, hence the order. Now we have
\begin{equation} 
\mathbb{E}\left( \sum_{i = 1}^n g_i^{L'}-H^{L'}\right)^2  = O(n^2L^22^{-L}).
\end{equation}

\setcounter{equation}{235}

For the third term, we apply the Cauchy--Schwarz inequality,
\begin{align}
& \quad 2\mathbb{E}\left(\sum_{i = 1}^n g_i^L - H^L\right)\left( \sum_{i = 1}^n g_i^{L'}-H^{L'}\right) \\
& \leq 2\sqrt{\mathbb{E}\left(\sum_{i = 1}^n g_i^L - H^L\right)^2}\sqrt{\mathbb{E}(\sum_{i = 1}^n g_i^{L'}-H^{L'})^2}  \\
& = O(n^{3/2}L^22^{-L/2})
\end{align}

Summing the three terms together and taking $L = 2\log n$, we have
\begin{equation}
\mathbb{E}\left|\sum_{i = 1}^n Z_{i}\right|^2 = O(n(\log n)^2)\label{bound.as}
\end{equation}
and thus
\begin{align}
\mathbb{E}\left|-\frac{1}{n}\log P(Y^n \| X^n) - \rH (\bY \| \bX)\right|
& = \mathbb{E}\left|\frac{1}{n}\sum_{i = 1}^n Z_i\right| \\
& \leq \frac{1}{n}\sqrt{\mathbb{E}\left|\sum_{i = 1}^n Z_{i}\right|^2} \\
& = O(n^{-1/2}\log n).
\end{align}

Now we deal with the almost sure convergence rates of AEP of causally conditional entropy rate.
We restate the G\'al--Koksma theorem\cite{Gal--Koksma1948} as follows:
\begin{lemma}[G\'al--Koksma theorem] Let $(\Omega, \mathcal{F},\mathbb{P})$ be a probability space and let $(Z_n)_{n\geq 1}$ be a sequence of random variables belonging to $L^p$, $p\geq 1$, such that
\begin{equation}
\mathbb{E}|Z_{M+1} + Z_{M+2} + \cdots + Z_{M+n}|^p = O(\Psi(n))
\end{equation}
uniformly in $M$, where $\Psi(n)/n$ is a nondecreasing sequence. Then for every $\epsilon>0$,
\begin{align}
&Z_1(\omega) + Z_2(\omega)+\cdots + Z_{n}(\omega) \nonumber\\
&\qquad\qquad = o((\Psi(n)(\log n)^{p+1+\epsilon})^{\frac{1}{p}})\quad \textrm{$P$-a.s.}
\end{align}
\end{lemma}

The bound in~(\ref{bound.as}) indicates that if we take $\Psi(n) = n(\log n)^2$ and $p = 2$ in
the G\'al--Koksma theorem, then for every $\epsilon>0$,
\begin{align}
-\frac{1}{n}\log P(Y^n\| X^n) - \rH(\bY \| \bX) & = o(n^{-1/2}(\log n)^{5/2 + \epsilon}) \\
& \quad  \quad \quad \quad \quad \textrm{$P$-a.s.}\label{gal.rate}
\end{align}

\subsection{Proof of Lemma~\ref{lemma.as}}
Denote the alphabet size $|\mathcal{X}|$ as $M$. We examine the updating computation of $P_w^\lambda(X_{n+1} = q|x^n)$, $q = 0,1,\ldots,M-1$. For an internal node $s$ in the updating path, if $js$ is in the updating path, we have~(\ref{eqn.weighting}). For the leaf node $v$ in the updating path,
\begin{equation}
P_w^v(X_{n+1} = q|x^n) = P_e^v(X_{n+1} = q|x^n).
\end{equation}
The computation of $P_w^\lambda(X_{n+1} = q|x^n)$ starts from a leaf and is repeated recursively along the updating path, until we reach the root node $\lambda$ and obtain $P_w^\lambda(X_{n+1} = q|x^n)$. Thus, $P_w^\lambda(X_{n+1} = q|x^n)$ is a weighted sum of $P_e^s(X_{n+1} = q|x^n)$, where $s$ is any node in the updating path.

Let $\{s\to \lambda\}$ denote the set of nodes in the path from $s$ to $\lambda$. The weight associated with $P_e^s(X_{n+1} = q|x^n)$ is
\begin{equation}\label{eqn.weightsinctw}
\beta^s(x^n)\prod_{u \in \{s\to \lambda\}} \frac{1}{\beta^u(x^n)+1},
\end{equation}
where $s$ is an internal node in the updating path. The weight associated with $P_w^v(X_{n+1} = q|x^n)$, where $v$ is the leaf node in the updating path, is
\begin{equation}\label{eqn.weightsinctw2}
\prod_{u\in \{\{u\to \lambda\}\backslash v\}} \frac{1}{\beta^u(x^n)+1}.
\end{equation}

The convergence properties of $P_w^\lambda (X_{n+1} = q|X^n)$ depends on the limiting behavior of $\beta^s(X^n)$ at every node $s$ along the updating path. If $s$ is an internal node in the tree representation of the source, we actually have $\lim_{n\to\infty} \beta^s(X^{n}) = 0$ almost surely. This fact was stated in \cite[Lemma 4]{Cai--Kulkarni--Verdu2006}. Here, we restate this fact and give a proof for stationary irreducible aperiodic finite-alphabet Markov processes.

\begin{lemma}\label{lemma.verdu}
Let $s$ be an internal node in the tree representation of the source. Then
\begin{equation}
\lim_{n\to\infty} \beta^s(X^{n}) = 0\quad \textrm{$P$-a.s.}
\end{equation}
\end{lemma}

\begin{IEEEproof}
It suffices to show
\begin{equation}
\lim_{n \to \infty}\frac{\beta^s(X^{n})}{\beta^s(X^{n})+1} = 0\quad \textrm{$P$-a.s.}
\end{equation}
We have
\begin{align}
&\quad \frac{\beta^s(X^{n})}{\beta^s(X^{n})+1} \\
& = \frac{P^s_e(X^{n})}{2P^s_w(X^{n})}  \\
& \leq \frac{P^s_e(X^{n})}{\prod_{i = 0}^{M-1} P^{is}_w(X^{n})} \\
& \leq 2^M \frac{P_e^s(X^n)}{\prod_{i = 0}^{M-1} P^{is}_e(X^{n})} \\
& = 2^M\exp\left \{ n_s \cdot \left( \frac{1}{n_s}\log P_e^s(X^n) - \frac{1}{n_s}\log\prod_{i = 0}^{M-1} P^{is}_e(X^{n})\right)\right \},
\end{align}
where $n_s$ denotes the number of symbols in $X^n$ with context $s$, and the inequalities follow from applying (\ref{eqn.howtoboundlemma}) repeatedly. Here since $s$ is an internal node of the tree, without loss of generality, we can assume offsprings of $s$ do not all have the same conditional distribution. If it were violated, we can simply iterate the inequalities obtain above till we reach the leaf nodes of the tree, after which we can apply the same arguments that will be shown later.

It was shown in \cite{Krichevsky--Trofimov1981} that the Krichevsky--Trofimov probability estimate of sequence $X^n$, i.e., $P_e(X^n)$, satisfies the following bound:
\begin{align}
& \quad \left | \frac{\log P_e(X^n)}{n} - \sum_{a\in \mathcal{X}}\frac{N(a|X^n)}{n}\log \frac{N(a|X^n)}{n} \right | \nonumber \\
& \leq \frac{M-1}{2}\frac{\log n}{n} + \frac{C}{n},\label{eqn.entropyconverge}
\end{align}
where $N(a|X^n)$ denotes the number of symbol $a$ in the sequence $X^n$, and $C$ is a constant depending only on the alphabet size $M$.

Under the assumption of Lemma~\ref{lemma.as}, Markov process $\mathbf{X}$ is ergodic, hence
\begin{equation}\label{eqn.convergetostat}
\lim_{n\to \infty}\frac{N(a|X^n)}{n} = \pi(a),
\end{equation}
where $\pi(\cdot)$ is the stationary distribution of $\mathbf{X}$. Equation~(\ref{eqn.convergetostat}) implies that
\begin{equation}
\lim_{n\to \infty}\frac{\log P_e(X^n)}{n} = -H(\pi).
\end{equation}

Applying the same argument to $\frac{1}{n_s}\log P_e^s(X^n)$, we have
\begin{equation}
\lim_{n\to \infty} \frac{1}{n_s}\log P_e^s(X^n) = -H(\pi_s),
\end{equation}
where $\pi_s$ is the stationary conditional distribution conditioned on context $s$. Analogously, for node $is$, we have
\begin{equation}
\lim_{n\to \infty} \frac{1}{n_{is}}\log P_e^{is}(X^n) = -H(\pi_{is}),
\end{equation}
thus
\begin{align}
& \quad \lim_{n\to \infty}\frac{1}{n_s}\log P_e^s(X^n) - \frac{1}{n_s}\log\prod_{i = 0}^{M-1} P^{is}_e(X^{n}) \nonumber\\
& =  -\left ( H(\pi_s) -\sum_{i \in \mathcal{X}} p_iH(\pi_{is}) \right),
\end{align}
where $p_i = P(\textrm{context is }is|\textrm{context is }s)$. It is obvious that
\begin{equation}
\pi_s = \sum_{i\in \mathcal{X}} p_i \pi_{is}.
\end{equation}

By the strict concavity of entropy functional and the fact that the offsprings of $s$ do not all have the same conditional distribution, we know
\begin{equation}
\lim_{n\to \infty}\frac{1}{n_s}\log P_e^s(X^n) - \frac{1}{n_s}\log\prod_{i = 0}^{M-1} P^{is}_e(X^{n}) <0,
\end{equation}
which implies
\begin{equation}
\lim_{n\to\infty}\frac{\beta^s(X^n)}{\beta^s(X^n) + 1} = 0\quad \textrm{$P$-a.s.}
\end{equation}
hence
\begin{equation}
\lim_{n\to\infty}\beta^s(X^n) = 0\quad \textrm{$P$-a.s.}
\end{equation}
holds.

\end{IEEEproof}

We know $Q(q|X^n) = P_w^\lambda(X_{n+1} = q|X^n)$ can be expressed as a weighted sum of $P_e^s(X_{n+1} = q|X^n)$ for $s$ in the updating path:
\begin{equation}
Q(q|X^n) = \sum_{s}w_sP_e^s(X_{n+1} = q|X^n),
\end{equation}
where $w_s$ are given in (\ref{eqn.weightsinctw}) and (\ref{eqn.weightsinctw2}). Lemma~\ref{lemma.verdu} implies that for $s$ an internal node of the tree representation of $\mathbf{X}$, $w_s \to 0$. Hence
\begin{equation}
\lim_{n\to \infty}Q(q|X^n) - \sum_{s\textrm{ leaf node}}w_sP_e^s(X_{n+1} = q|X^n) = 0\quad \textrm{$P$-a.s.}
\end{equation}
For leaf node $s$, by the property of Krichevsky--Trofimov probability estimate, we know
\begin{equation}
\lim_{n\to \infty}P_e^s(X_{n+1} = q|X^n)-P(q|X^n) = 0,
\end{equation}
where $P(q|X^n)$ is the true conditional probability. Thus we have
\begin{align}
& Q(x_{n+1}|X^{n}) - P(x_{n+1}|X^{n}) \nonumber\\
&\qquad = P^{\lambda}_w(x_{n+1}|X^{n}) - P(x_{n+1}|X^{n})  \\
&\qquad \to 0\quad\textrm{$P$-a.s.}
\end{align}

\subsection{Proof of Lemma~\ref{lemma.unifconv}}
Fix $\epsilon >0$. Since $\mathcal{M}(\cX,\cY)$ is bounded and closed, $f(\cdot)$ is uniformly continuous. Thus there exists $\delta_\epsilon$ such that $|f(P)-f(Q)|\leq \epsilon$ if $\norm{P-Q}_1\leq \delta_\epsilon$. Furthermore, $f(\cdot)$ is bounded by $f_\mathrm{max}\triangleq\log |\cX|+\log |\cY|$. Therefore, we have
\begin{align}
|f(P)-f(Q)|
&\leq \epsilon \mathbf{1}_{\{\norm{P-Q}_1\leq \delta_\epsilon\}}+ f_\mathrm{max} \mathbf{1}_{\{\norm{P-Q}_1 > \delta_\epsilon\}}\\
            &\leq \epsilon + f_\mathrm{max} \frac{\norm{P-Q}_1}{\delta_\epsilon}\\
            &=\epsilon + K_\epsilon \norm{P-Q}_1,
\end{align}
where $K_\epsilon =  f_\mathrm{max}/\delta_\epsilon$.

\subsection{Proof of Lemma~\ref{lemma.diff}}
Since
\begin{equation}
H(Y|X) = H(X,Y) - H(X),
\end{equation}
we can bound $|f(P)-f(Q)|$ as
\begin{align}
&|f(P)-f(Q)| \nonumber\\
& = |H_P(X,Y)-H_P(X)-H_Q(X,Y)+H_Q(X)| \\
&\leq |H_P(X,Y)-H_Q(X,Y)|+|H_P(X)-H_Q(X)|.
\end{align}
Now, by \cite[Lemma 2.7]{Csiszar--Korner1981}, we have
\begin{align}
|H_P(X,Y)-H_Q(X,Y)| & \leq \theta \log \frac{|\mathcal{X}||\mathcal{Y}|}{\theta}, \\
|H_P(X)-H_Q(X)| & \leq \theta_X \log \frac{|\mathcal{X}|}{\theta_X},
\end{align}
where $\theta = \| P_{XY}-Q_{XY} \|_1$ and $\theta_X = \| P_{X} - Q_{X} \|_1$. Since
\begin{align}
\theta & = \sum_{x\in \mathcal{X},y\in \mathcal{Y}}|P(x,y) - Q(x,y)| \\
& = \sum_{x\in \mathcal{X}}\sum_{y\in \mathcal{Y}} |P(x,y) - Q(x,y)| \\
& \geq \sum_{x\in \mathcal{X}}\left |\sum_{y\in \mathcal{Y}} P(x,y) - Q(x,y) \right| \\
&  =   \sum_{x\in \mathcal{X}} |P(x) - Q(x)| \\
& = \theta_X,
\end{align}
we have
\begin{equation}
|f(P)-f(Q)| \leq 2\theta \log \frac{|\mathcal{X}||\mathcal{Y}|}{\theta}.
\end{equation}

\subsection{Proof of Lemma~\ref{lemma.mixing}}
We first define the $\alpha$-mixing coefficient of a stationary process.
\begin{definition}[$\alpha$-mixing coefficient]
For a stationary process $\mathbf{X}$ adapted to the filtration $(\mathcal{F}_n)_{-\infty}^{\infty}$, the $\alpha$-mixing coefficient is defined as
\begin{equation}
\alpha(n) \triangleq \sup|P(A\cap B)-P(A)P(B)|,
\end{equation}
where the supremum is over
all $A\in \mathcal{F}_{-\infty}^0$ and $B\in \mathcal{F}_n^{\infty}$.
\end{definition}
According to \cite{Bradley2005}, if $\mathbf{X}$ is a stationary irreducible aperiodic Markov process, $\alpha(n)$ tends to zero exponentially fast in $n$, i.e., there exist
$C_7>0$ and $C_8>0$ such that
\begin{equation}\label{eqn.alphamixingbound}
\alpha(n)\leq C_7e^{-C_8n}.
\end{equation}

We bound $\mathbb{E}\left ((1/n) \sum_{i = 1}^n V_i \right )^2$ as follows:
\begin{align}
\mathbb{E}\left|\frac{1}{n} \sum_{i = 1}^n V_i\right|^2
 & = \frac{1}{n^2} \sum_{i = 1}^n \mathbb{E}|V_i|^2 + \frac{2}{n^2} \sum_{1\leq i<j\leq n} \mathbb{E}V_iV_j \\ \label{eqn.boundedbyv}
 & \leq \frac{V^2}{n} + \frac{2}{n^2} \sum_{1\leq i<j\leq n} \mathbb{E}V_iV_j,
\end{align}
where (\ref{eqn.boundedbyv}) holds because $V_i, \forall i$ is uniformly bounded by constant $V$.

By Billingsley's inequality\cite[Corollary 1.1]{Bosq1996}, taking into account that $\mathbb{E}V_i = 0, \forall i$, we know the following bound holds:
\begin{equation}\label{eqn.bosqbillingsley}
|\mathbb{E}V_iV_j| = |\mathsf{Cov}(V_i,V_j)| \leq 4 V^2 \alpha(|i-j|).
\end{equation}

Plugging (\ref{eqn.bosqbillingsley}) into (\ref{eqn.boundedbyv}), we have
\begin{align}
\mathbb{E}\left|\frac{1}{n} \sum_{i = 1}^n V_i\right|^2 & \leq \frac{V^2}{n} + \frac{8V^2}{n^2}\sum_{1\leq i<j\leq n} \alpha(|i-j|)\\
& \leq \frac{V^2}{n} + \frac{8V^2}{n^2}C_7\sum_{k = 1}^{n-1}ke^{-C_8(n-k)} \\
& \leq \frac{V^2}{n} + \frac{8C_7V^2e^{C_8}}{n(e^{C_8}-1)^2},
\end{align}

Thus, we show Lemma~\ref{lemma.mixing} holds with $C_4 = 1+8C_7e^{C_8}/(e^{C_8}-1)^2$.

\bibliographystyle{IEEEtran}
\bibliography{di}

\newcommand{\noopsort}[1]{}
\begin{thebibliography}{10}
\providecommand{\url}[1]{#1}
\csname url@samestyle\endcsname
\providecommand{\newblock}{\relax}
\providecommand{\bibinfo}[2]{#2}
\providecommand{\BIBentrySTDinterwordspacing}{\spaceskip=0pt\relax}
\providecommand{\BIBentryALTinterwordstretchfactor}{4}
\providecommand{\BIBentryALTinterwordspacing}{\spaceskip=\fontdimen2\font plus
\BIBentryALTinterwordstretchfactor\fontdimen3\font minus
  \fontdimen4\font\relax}
\providecommand{\BIBforeignlanguage}[2]{{%
\expandafter\ifx\csname l@#1\endcsname\relax
\typeout{** WARNING: IEEEtran.bst: No hyphenation pattern has been}%
\typeout{** loaded for the language `#1'. Using the pattern for}%
\typeout{** the default language instead.}%
\else
\language=\csname l@#1\endcsname
\fi
#2}}
\providecommand{\BIBdecl}{\relax}
\BIBdecl

\bibitem{Marko1973}
H.~Marko, ``The bidirectional communication theory--a generalization of
  information theory,'' \emph{{IEEE} Trans. Commum.}, vol. COM-21, pp.
  1345--1351, 1973.

\bibitem{Massey90}
J.~L. Massey, ``Causality, feedback, and directed information,'' in \emph{Proc.
  Int. Symp. Inf. Theory Appl.}, Honolulu, HI, Nov. 1990, pp. 303--305.

\bibitem{Kramer1998}
G.~Kramer, \emph{Directed Information for Channels with Feedback}.\hskip 1em
  plus 0.5em minus 0.4em\relax Konstanz: Hartung-Gorre Verlag, 1998, {D}r. sc.
  thchn. Dissertation, Swiss Federal Institute of Technology (ETH) Zurich.

\bibitem{Kramer03}
------, ``Capacity results for the discrete memoryless network,'' \emph{{IEEE}
  Trans. Inf. Theory}, vol.~49, no.~1, pp. 4--21, 2003.

\bibitem{Tatikonda--Sanjoy2009}
S.~Tatikonda and S.~Mitter, ``The capacity of channels with feedback,''
  \emph{{IEEE} Trans. Inf. Theory}, vol.~55, no.~1, pp. 323--349, 2009.

\bibitem{Kim08}
Y.-H. Kim, ``A coding theorem for a class of stationary channels with
  feedback,'' \emph{{IEEE} Trans. Inf. Theory}, vol.~54, no.~4, pp. 1488--1499,
  2008.

\bibitem{Permuter--Weissman--Goldsmith2009}
H.~H. Permuter, T.~Weissman, and A.~J. Goldsmith, ``Finite state channels with
  time-invariant deterministic feedback,'' \emph{{IEEE} Trans. Inf. Theory},
  vol.~55, no.~2, pp. 644--662, 2009.

\bibitem{Permuter--Kim--Weissman2009}
H.~H. Permuter, Y.-H. Kim, and T.~Weissman, ``Interpretations of directed
  information in portfolio theory, data compression, and hypothesis testing,''
  \emph{{IEEE} Trans. Inf. Theory}, vol.~57, no.~3, pp. 3248--3259, Jun. 2011.

\bibitem{Granger1969}
C.~Granger, ``Investigating causal relations by econometric models and
  cross-spectral methods,'' \emph{Econometrica}, vol.~37, no.~3, pp. 424--438,
  1969.

\bibitem{Mathai--Martins--Shapiro2007}
P.~Mathai, N.~C. Martins, and B.~Shapiro, ``On the detection of gene network
  interconnections using directed mutual information,'' in \emph{Proc. {UCSD}
  Inf. Theory Appl. Workshop}, 2007.

\bibitem{Rao--Hero--States--Engel2008}
A.~Rao, A.~O. Hero, D.~J. States, and J.~D. Engel, ``Using directed information
  to build biologically relevant influence networks,'' \emph{J. Bioinform.
  Comput. Biol.}, vol.~6, no.~3, pp. 493--519, 2008.

\bibitem{Verdu2005}
S.~Verd{\'u}, ``Universal estimation of information measures,'' in \emph{Proc.
  {IEEE} Inf. Theory Workshop}, 2005.

\bibitem{Wyner--Ziv1989}
A.~D. Wyner and J.~Ziv, ``Some asymptotic properties of the entropy of a
  stationary ergodic data source with applications to data compression,''
  \emph{{IEEE} Trans. Inf. Theory}, vol.~35, no.~6, pp. 1250--1258, 1989.

\bibitem{Ziv--Merhav1993}
J.~Ziv and N.~Merhav, ``A measure of relative entropy between individual
  sequences with application to universal classification,'' \emph{{IEEE} Trans.
  Inf. Theory}, vol.~39, no.~4, pp. 1270--1279, 1993.

\bibitem{Cai--Kulkarni--Verdu2006}
H.~Cai, S.~R. Kulkarni, and S.~Verd{\'u}, ``Universal divergence estimation for
  finite-alphabet sources,'' \emph{{IEEE} Trans. Inf. Theory}, vol.~52, no.~8,
  pp. 3456--3475, 2006.

\bibitem{Burrows--Wheeler1994}
M.~Burrows and D.~J. Wheeler, \emph{A block-sorting lossless data compression
  algorithm}.\hskip 1em plus 0.5em minus 0.4em\relax Digital Systems Research
  Center, Tech. Rep. 124, 1994.

\bibitem{Willems--Shtarkov--Tjalkens1995}
F.~M.~J. Willems, Y.~M. Shtarkov, and T.~J. Tjalkens, ``The context-tree
  weighting method: Basic properties,'' \emph{{IEEE} Trans. Inf. Theory},
  vol.~41, no.~3, pp. 653--664, 1995.

\bibitem{Cai--Kulkarni--Verdu2004}
H.~Cai, S.~R. Kulkarni, and S.~Verd{\'u}, ``Universal entropy estimation via
  block sorting,'' \emph{{IEEE} Trans. Inf. Theory}, vol.~50, no.~7, pp.
  1551--1561, 2004.

\bibitem{Yu--Verdu2006}
J.~Yu and S.~Verd{\'u}, ``Universal erasure entropy estimation,'' in
  \emph{Proc. {IEEE} Int. Symp. Inf. Theory}, 2006.

\bibitem{Quinn--Coleman--Kiyavash--Hatsopoulos2009}
C.~J. Quinn, T.~P. Coleman, N.~Kiyavash, and N.~G. Hatsopoulos, ``Estimating
  the directed information to infer causal relationships in ensemble neural
  spike train recordings,'' \emph{J. Comput. Neurosci.}, 2011.

\bibitem{Zhao--Kim--Permuter--Weissman2010}
L.~Zhao, Y.-H. Kim, H.~H. Permuter, and T.~Weissman, ``Universal estimation of
  directed information,'' in \emph{Proc. {IEEE} Int. Symp. Inf. Theory}, 2010,
  pp. 230--234.

\bibitem{Massey--Massey2005}
J.~L. Massey and P.~C. Massey, ``Conservation of mutual and directed
  information,'' in \emph{Proc. {IEEE} Int. Symp. Inf. Theory}, 2005, pp.
  157--158.

\bibitem{Amblard--Michel2011}
\BIBentryALTinterwordspacing
P.-O. Amblard and O.~J.~J. Michel, ``Relating {G}ranger causality to directed
  information theory for networks of stochastic processes,'' 2011. [Online].
  Available: \url{http://arxiv.org/abs/0911.2873v4}
\BIBentrySTDinterwordspacing

\bibitem{Cover--Thomas2006}
T.~M. Cover and J.~A. Thomas, \emph{Elements of Information Theory},
  2nd~ed.\hskip 1em plus 0.5em minus 0.4em\relax New York: Wiley, 2006.

\bibitem{Ornstein1978}
D.~Ornstein, ``Guessing the next output of a stationary process,'' \emph{Israel
  J. Math.}, vol.~30, pp. 292--296, 1978.

\bibitem{Algoet1992}
P.~Algoet, ``Universal schemes for prediction, gambling and portfolio
  selection,'' \emph{Ann. Prob.}, vol.~20, pp. 901--941, 1992.

\bibitem{Morvai--Yakowitz--Algoet1997}
G.~Morvai, S.~J. Yakowitz, and P.~Algoet, ``Weakly convergent nonparametric
  forecasting of stationary time series,'' \emph{{IEEE} Trans. Inf. Theory},
  vol.~43, no.~2, pp. 483--498, 1997.

\bibitem{MerhavFeder98}
N.~Merhav and M.~Feder, ``Universal prediction,'' \emph{{IEEE} Trans. Inf.
  Theory}, vol.~44, no.~6, pp. 2124--2147, 1998.

\bibitem{Willems--Tjalkens1997}
F.~Willems and T.~Tjalkens, \emph{Complexity Reduction of the Context-Tree
  Weighting Algorithm: A Study for KPN Research}.\hskip 1em plus 0.5em minus
  0.4em\relax Tech. Rep. Univ. Eindhoven, Eindhoven, The Netherlands, EIDMA
  Rep. RS.97.01, 1997.

\bibitem{Tjalkens--Shtarkov--Willems1993}
T.~J. Tjalkens, Y.~M. Shtarkov, and F.~M.~J. Willems, ``Sequential weighting
  algorithms for multi-alphabet sources,'' in \emph{6th Joint Swedish--Russian
  Int. Workshop Inf. Theory}, 1993, pp. 230--234.

\bibitem{Willems1998}
F.~M.~J. Willems, ``The context-tree weighting method: Extensions,''
  \emph{{IEEE} Trans. Inf. Theory}, vol.~44, no.~2, pp. 792--798, 1998.

\bibitem{Krichevsky--Trofimov1981}
R.~E. Krichevsky and V.~K. Trofimov, ``The performance of universal encoding,''
  \emph{{IEEE} Trans. Inf. Theory}, vol.~27, no.~2, pp. 199--207, 1981.

\bibitem{Venkataramanan--Pradhan2007}
R.~Venkataramanan and S.~S. Pradhan, ``Source coding with feed-forward:
  {R}ate--distortion theorems and error exponents for a general source,''
  \emph{{IEEE} Trans. Inf. Theory}, vol.~53, no.~6, pp. 2154--2179, 2007.

\bibitem{Birch1962}
J.~Birch, ``Approximations for the entropy for functions of markov chains,''
  \emph{Ann. Math. Statist.}, vol.~33, pp. 930--938, 1962.

\bibitem{Gland--Mevel2000}
F.~L. Gland and L.~Mevel, ``Exponential forgetting and geometric ergodicity in
  hidden markov models,'' \emph{Math. Control Signals Syst.}, vol.~13, no.~1,
  pp. 63--93, 2000.

\bibitem{Hochwald--Jelenkovic1999}
B.~M. Hochwald and P.~Jelenkovi\'c, ``State learning and mixing in entropy of
  hidden {M}arkov processes and the {G}ilbert--{E}lliott channel,''
  \emph{{IEEE} Trans. Inf. Theory}, vol.~45, no.~1, pp. 128--138, 1999.

\bibitem{Kleinberg--Hripcsak2011}
S.~Kleinberg and G.~Hripcsak, ``A review of causal inference for biomedical
  informatics,'' \emph{J. Biomed. Inform.}, vol.~44, no.~6, pp. 1102--1112,
  2011.

\bibitem{Breiman1957}
L.~Breiman, ``The individual ergodic theorem of information theory,''
  \emph{Ann. Math. Statist.}, vol.~28, no.~3, pp. 809--811, 1957, correction
  (1960). 31(3), 809--810.

\bibitem{Pinsker1964}
M.~S. Pinsker, \emph{Information and Information Stability of Random Variables
  and Processes}.\hskip 1em plus 0.5em minus 0.4em\relax San Francisco:
  Holden-Day, 1964.

\bibitem{Barron1986}
A.~R. Barron, ``Entropy and the central limit theorem,'' \emph{Ann. Probab.},
  vol.~14, pp. 336--342, 1986.

\bibitem{Breiman1992}
L.~Breiman, \emph{Probability}.\hskip 1em plus 0.5em minus 0.4em\relax SIAM:
  Society for Industrial and Applied Mathematics, 1992.

\bibitem{Tsybakov2008}
A.~Tsybakov, \emph{Introduction to Nonparametric Estimation}.\hskip 1em plus
  0.5em minus 0.4em\relax Springer-Verlag, 2008.

\bibitem{McMillan1953}
B.~McMillan, ``The basic theorems of information theory,'' \emph{Ann. Math.
  Statist.}, vol.~24, no.~2, pp. 196--219, 1953.

\bibitem{Gal--Koksma1948}
I.~S. G{\'a}l and J.~F. Koksma, ``Sur l'ordre de grandeur des fonctions
  sommables,'' \emph{C. R. Acad. Sci. Paris}, vol. 227, pp. 1321--1323, 1948.

\bibitem{Csiszar--Korner1981}
I.~Csisz{\'a}r and J.~K{\"o}rner, \emph{\noopsort{c}{I}nformation Theory:
  Coding Theorems for Discrete Memoryless Systems}.\hskip 1em plus 0.5em minus
  0.4em\relax Budapest: Akad\'emiai Kiad\'o, 1981.

\bibitem{Bradley2005}
R.~Bradley, ``Basic properties of strong mixing conditions. a survey and some
  open questions,'' \emph{Probab. Surveys}, vol.~2, pp. 107--144, 2005.

\bibitem{Bosq1996}
D.~Bosq, ``Nonparametric statistics for stochastic processes,'' \emph{Lecture
  Notes in Statist}, 1996.

\end{thebibliography}

\begin{IEEEbiographynophoto}{Jiantao Jiao}
(SM'13) received the B.Eng. degree with the highest honor in Electronic Engineering from Tsinghua University, Beijing, China, in 2012. He is currently working towards the Ph.D. degree in the Department of Electrical Engineering, Stanford University. His research interests include information theory and statistical signal processing, with applications in communication, control,
computation, networking, data compression, and learning.

Mr. Jiao is a recipient of the Stanford Graduate Fellowship (SGF), the highest award offered by Stanford University.
\end{IEEEbiographynophoto}

\begin{IEEEbiographynophoto}{Haim Permuter}
(M'08) received his B.Sc.\@ (summa cum laude) and M.Sc.\@ (summa cum laude) degrees in Electrical and
Computer Engineering from the Ben-Gurion University, Israel, in 1997 and 2003, respectively, and the
Ph.D. degree in Electrical Engineering from Stanford University, California in 2008.

Between 1997 and 2004, he was an officer at a research and development unit of the Israeli
Defense Forces. He is currently a senior lecturer at Ben-Gurion university.

Dr. Permuter is a recipient of the Fullbright Fellowship, the Stanford Graduate Fellowship (SGF),
Allon Fellowship, and and the 2009 U.S.-Israel Binational Science Foundation Bergmann Memorial
Award.\end{IEEEbiographynophoto}

\begin{IEEEbiographynophoto}{Lei Zhao}
received the B.Eng. degree from Tsinghua University, China, in 2003, the M.S. degree in Electrical and Computer Engineering from Iowa State University, Ames, in 2006, and the Ph.D. degree in Electrical Engineering from Stanford University, California in 2011.

Dr. Zhao is currently working at Jump Operations, Chicago, IL, USA.
\end{IEEEbiographynophoto}

\begin{IEEEbiographynophoto}{Young-Han Kim}
(S'99--M'06--SM'12) received the B.S. degree with honors in electrical engineering from Seoul National
University, Seoul, Korea, in 1996 and the M.S. degrees in electrical engineering and in statistics, and the
Ph.D. degree in electrical engineering from Stanford University, Stanford, CA, in 2001, 2006, and
2006, respectively.

In July 2006, he joined the University of California, San Diego, where he is an Associate
Professor of Electrical and Computer Engineering. His research interests are in statistical
signal processing and information theory, with applications in communication, control,
computation, networking, data compression, and learning.

Dr. Kim is a recipient of the 2008 NSF Faculty Early Career Development (CAREER) Award the 2009 US-Israel Binational Science Foundation Bergmann Memorial Award, and the 2012 IEEE Information Theory Paper Award. He is currently on the Editorial Board of the
\textsc{IEEE Transactions on Information Theory,} serving as an Associate Editor for Shannon theory. He is also serving as a Distinguished Lecturer for the IEEE Information Theory Society.\end{IEEEbiographynophoto}

\begin{IEEEbiographynophoto}{Tsachy Weissman}
(S'99-M'02-SM'07-F'13) graduated summa cum laude with a
B.Sc. in electrical engineering from the Technion in 1997, and earned
his Ph.D. at the same place in 2001. He then worked at Hewlett-Packard
Laboratories with the information theory group until 2003, when he joined
Stanford University, where he is Associate Professor of Electrical
Engineering and incumbent of the
STMicroelectronics chair in the School of Engineering.
He has spent leaves at the Technion, and at ETH Zurich.

Tsachy's research is focused on information theory, statistical signal
processing, the interplay between them, and their applications.

He is recipient of several best paper awards, and prizes for excellence in research.

He currently serves on the editorial boards of the \textsc{IEEE Transactions on Information Theory} and Foundations and Trends in Communications and
Information Theory.
\end{IEEEbiographynophoto}

\end{document}